\documentclass[a4paper,twocolumn,11pt,accepted=2024-02-05]{quantumarticle}
\pdfoutput=1
\usepackage[numbers,sort&compress]{natbib}
\usepackage[utf8]{inputenc}
\usepackage[english]{babel}
\usepackage[T1]{fontenc}

\usepackage{soul}
\usepackage{graphicx}
\usepackage{amssymb,amsmath,color,amsfonts, amsthm}
\usepackage{hyperref}
\hypersetup{
	colorlinks,
	linkcolor={blue},
	citecolor={blue},
	urlcolor={blue}
}
\usepackage{ragged2e}
\usepackage[capitalise]{cleveref}
\usepackage[normalem]{ulem}
\usepackage[utf8]{inputenc}
\usepackage[english]{babel}
\usepackage{units}
\usepackage{braket}
\usepackage{float}
\usepackage{epstopdf}
\newcommand{\suptiny}[3]{\ensuremath{^{\hspace{#1 pt}\protect\raisebox{#2 pt}{\tiny{$ #3$}}}}}

\begin{document}

\title{Photonic entanglement during a zero-g flight}

\author{Julius Arthur Bittermann}
\orcid{0000-0003-1950-106X}
\email{Julius.Bittermann@oeaw.ac.at}
\affiliation{Institute for Quantum Optics and Quantum Information -- IQOQI Vienna, Austrian Academy of Sciences, Boltzmanngasse 3, 1090 Vienna, Austria}
\affiliation{Atominstitut,  Technische  Universit{\"a}t  Wien,  Stadionallee 2, 1020  Vienna,  Austria}
\author{Lukas Bulla}
\orcid{0000-0002-7978-3110}
\affiliation{Institute for Quantum Optics and Quantum Information -- IQOQI Vienna, Austrian Academy of Sciences, Boltzmanngasse 3, 1090 Vienna, Austria}
\affiliation{present address: Quantum Technology Laboratories GmbH, Clemens-Holzmeister-Stra{\ss}e 6/6, 1100 Vienna, Austria}
\author{Sebastian Ecker}
\orcid{0000-0002-7597-741X}
\affiliation{Institute for Quantum Optics and Quantum Information -- IQOQI Vienna, Austrian Academy of Sciences, Boltzmanngasse 3, 1090 Vienna, Austria}
\affiliation{present address: Quantum Technology Laboratories GmbH, Clemens-Holzmeister-Stra{\ss}e 6/6, 1100 Vienna, Austria}
\author{Sebastian Philipp Neumann}
\orcid{0000-0002-5968-5492}
\affiliation{Institute for Quantum Optics and Quantum Information -- IQOQI Vienna, Austrian Academy of Sciences, Boltzmanngasse 3, 1090 Vienna, Austria}
\affiliation{present address: Quantum Technology Laboratories GmbH, Clemens-Holzmeister-Stra{\ss}e 6/6, 1100 Vienna, Austria}
\author{Matthias Fink}
\orcid{0000-0001-7459-4508}
\affiliation{Institute for Quantum Optics and Quantum Information -- IQOQI Vienna, Austrian Academy of Sciences, Boltzmanngasse 3, 1090 Vienna, Austria}
\affiliation{present address: Quantum Technology Laboratories GmbH, Clemens-Holzmeister-Stra{\ss}e 6/6, 1100 Vienna, Austria}
\author{Martin Bohmann}
\orcid{0000-0003-3857-4555}
\email{martin.bohmann@qtlabs.at}
\affiliation{Institute for Quantum Optics and Quantum Information -- IQOQI Vienna, Austrian Academy of Sciences, Boltzmanngasse 3, 1090 Vienna, Austria}
\affiliation{present address: Quantum Technology Laboratories GmbH, Clemens-Holzmeister-Stra{\ss}e 6/6, 1100 Vienna, Austria}
\author{Nicolai Friis}
\orcid{0000-0003-1950-8640}
\email{nicolai.friis@tuwien.ac.at}
\affiliation{Atominstitut,  Technische  Universit{\"a}t  Wien,  Stadionallee 2, 1020  Vienna,  Austria}
\affiliation{Institute for Quantum Optics and Quantum Information -- IQOQI Vienna, Austrian Academy of Sciences, Boltzmanngasse 3, 1090 Vienna, Austria}
\author{Marcus Huber}
\orcid{0000-0003-1985-4623}
\email{marcus.huber@tuwien.ac.at}
\affiliation{Atominstitut,  Technische  Universit{\"a}t  Wien,  Stadionallee 2, 1020  Vienna,  Austria}
\affiliation{Institute for Quantum Optics and Quantum Information -- IQOQI Vienna, Austrian Academy of Sciences, Boltzmanngasse 3, 1090 Vienna, Austria}
\author{Rupert Ursin}
\orcid{0000-0002-9403-269X}
\email{rupert.ursin@qtlabs.at}
\affiliation{Institute for Quantum Optics and Quantum Information -- IQOQI Vienna, Austrian Academy of Sciences, Boltzmanngasse 3, 1090 Vienna, Austria}
\affiliation{present address: Quantum Technology Laboratories GmbH, Clemens-Holzmeister-Stra{\ss}e 6/6, 1100 Vienna, Austria}


\begin{abstract}
Quantum technologies have matured to the point that we can test fundamental quantum phenomena under extreme conditions. Specifically, entanglement, a cornerstone of modern quantum information theory, can be robustly produced and verified in various adverse environments. We take these tests further and implement a high-quality Bell experiment during a parabolic flight, transitioning from microgravity to hypergravity of 1.8 g while continuously
observing Bell violation, with Bell-CHSH parameters between $S=-2.6202$
and $-2.7323$, an average of
$\overline{S} = -2.680$, and average standard deviation of $\overline{\Delta S} = 0.014$.
This violation is unaffected both by uniform and non-uniform acceleration.
This experiment demonstrates the stability of current quantum communication platforms for space-based applications and adds an important reference point for testing the interplay of non-inertial motion and quantum information.
\end{abstract}


\maketitle

\vspace*{1mm}

\noindent\textbf{Introduction}.\ Entanglement was once seen as a peculiar feature, relevant `only' to foundational questions that were considered difficult, if not impossible to conclusively test in experiments. And indeed, first milestone experiments in the 70s, 80s, and 90s demonstrating the utility of entanglement for Bell-inequality violation~\cite{FreedmanClauser1972,AspectGrangierRoger1981,AspectGrangierRoger1982,AspectDalibardRoger1982,WeihsJenneweinSimonWeinfurterZeilinger1998} \textemdash\ a task that is today considered to be a main primitive of quantum-information processing \textemdash\ had to overcome severe practical challenges. These efforts were recognized by the 2022 Nobel Prize in Physics. Recent decades have seen steady progress in building setups demonstrating entanglement, which culminated in closing all relevant loopholes for local-realist explanations of the observed correlations~\cite{Shalm-etal-loophole-free2015,Hensen-etal-loophole-free2015,Giustina-Zeilinger-loophole-free2015}. Nowadays, the generation and verification of entanglement under idealized laboratory conditions, even between multiple parties~\cite{FriisMartyEtal2018,Gong-Pan2019a,PogorelovEtAl2021,MooneyWhiteHillHollenberg2021b} and in high dimensions~\cite{WangEtAl2018,BavarescoEtAl2018,SchneelochTisonFantoAlsingHowland2019,Herrera-ValenciaSrivastavPivoluskaHuberFriisMcCutcheonMalik2020}, has become routine for a variety of platforms~\cite{FriisVitaglianoMalikHuber2019}.

A pertinent concern for the future exploration of entanglement-based quantum-communication technologies thus is the robustness of entanglement and of its applications under non-ideal, real-world conditions. On the one hand, experiments in this direction are driven by the curiosity of determining the practical limitations of this technology. For instance, one may ask how much noise and disturbance a setup can tolerate in principle while still operating within desired specifications (e.g., as in~\cite{Ecker-Huber2019}). On the other hand, the exquisite control over entanglement generating setups allows us to precisely identify or ultimately to rule out potentially detrimental effects and thus to learn more about the physical environments in which quantum-information processing protocols are carried out. Here we present a test of entanglement that incorporates both of these aspects.

We report on an experiment to test the violation of the Clauser-Horne-Shimony-Holt (CHSH) inequality~\cite{ClauserHorneShimonyHolt1969} by polarization-entangled photon pairs during a parabolic flight on a (modified) commercial airliner.
To this end, we built a compact laboratory necessary for the generation and measurement of entangled photons and installed it into an Airbus A310 of the company Novespace operating out of Bordeaux. In order to certify entanglement, we measured the Bell-CHSH parameter $S$ and the visibilities for accelerations in a range between $1.8$~g and microgravity, and expose the entangled system to continuous transitions from hyper- to microgravity and back from micro- to hypergravity, with uninterrupted phases of microgravity of about 20~s in between.

The results of this test show that the setup is stable throughout the flight manoeuvres and accompanying environmental changes (such as air pressure and temperature).
In particular, the statistics of the Bell-CHSH parameter do not allow us to distinguish between periods of micro- and hypergravity, or the periods of changing accelerations in between, and neither do similar tests for the measured visibilities during steady flight and microgravity. Our experiment thus demonstrates the remarkable tolerance of state-of-the-art sources of entangled photons to adverse environmental conditions, and it confirms with high precision that entanglement-based setups for quantum communication operate reliably across a range of gravitational regimes spanning three orders of magnitude.

\begin{figure*}[ht!]
\centering
\includegraphics[width=0.9\textwidth]{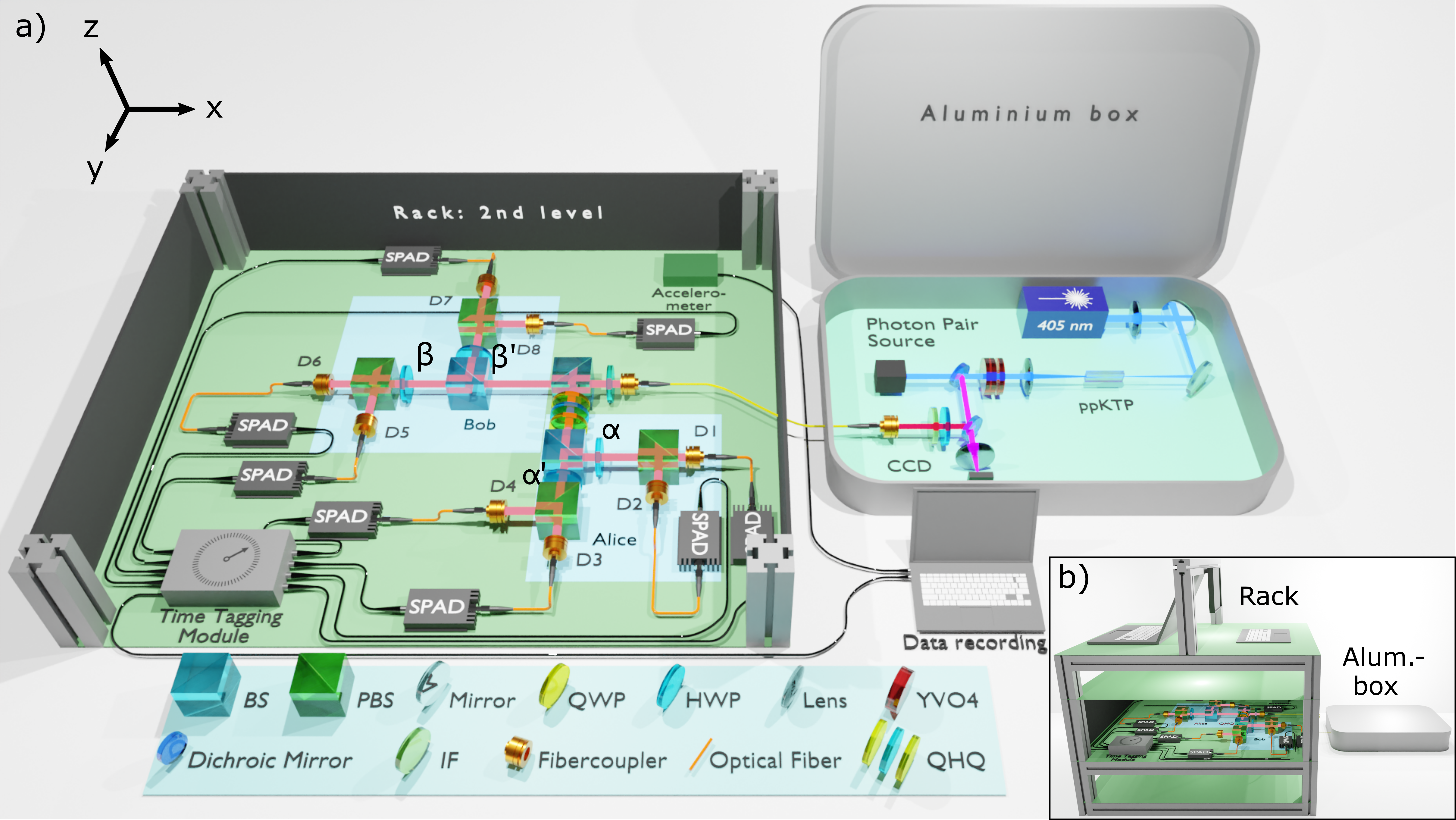}
\caption{\textbf{Illustration of the setup for measuring polarization entanglement during a zero-g flight. a)} Optical setup for the generation and certification of polarization entanglement, made up of a photon-pair source and two detection modules of Alice and Bob. In the source, the beam of a continuous-wave 405-nm pump laser diode is focused into a periodically poled type-II KTiOPO$_{4}$ (ppKTP) crystal. Through spontaneous parametric down-conversion a pump photon is converted into two orthogonally polarized photons at 810 nm. Two dichroic mirrors and an interference filter (IF) separate the down-converted photons from the laser beam, which are finally coupled into a polarization-maintaining single-mode fiber (PMSMF). The PMSMF guides the entangled photons to a 50:50 beamsplitter (BS), where the photon pair is spatially separated.
Four polarization analysis modules, each consisting of a half-wave plate (HWP) and a polarizing beamsplitter (PBS) with two single-photon avalanche diodes (SPAD) at the outputs, enable simultaneous polarization measurements of the four polarization angles $\alpha$, $\alpha'$, $\beta$ and $\beta'$ necessary for a CHSH-type Bell test (without having to change the settings in between).
D1-D8 label the detector numbers for the data analysis.
A time-tagging module and a computer record the arrival times of the photons and identify coincidences. An accelerometer measures the accelerations along the $x$, $y$, and $z$ axes. We use a quarter-, half-, and quarter-wave plate (QHQ) arrangement in one branch (Alice) to adjust the phase of the Bell state.
\textbf{b)} An aluminium box (600 mm $\times$ 400 mm $\times$ 250 mm) contains the photon-pair source, a rack (745 mm $\times$ 545 mm $\times$ 1230 mm) made up of strut profiles and aluminium plates contains the detection modules of Alice and Bob and the measurement devices. Rack and box are attached to the floor of the aircraft cabin. A photograph of the setup installed in the aircraft cabin can be found in Fig.~\ref{fig:setup_in_aircraft} in the appendix.}
\label{fig:setup}
\end{figure*}

In the wake of recent robustness tests for quantum-optical systems~\cite{Fink,MiciusI,MiciusII,HOM_Rot}, our experiment thus sets new standards for the durability of quantum-communication setups and their ability to generate and detect entanglement.
And even though the involved accelerations are far away from regimes where one would expect parametric amplification effects (e.g., the dynamical Casimir effect and related phenomena), our results provide a reference point for future studies of kinematic effects and non-uniform motion in quantum information.\\[-2mm]

\noindent\textbf{Experimental setup}.\ We employ an ultra-compact source for photon-pair generation embedded in an aluminium box, sketched in Fig.~\ref{fig:setup}.
A 10 mm long periodically poled type-II KTP crystal is placed in the focus of a continuous-wave 405~nm pump laser.
Pump photons are converted into orthogonally polarized spectrally degenerate photon pairs at 810~nm via spontaneous parametric down-conversion (SPDC). The birefringence in the ppKTP crystal causes a relative delay of the $H$ and $V$ photons, that we compensate using
two YVO$_{4}$ plates and a polarization-maintaining single-mode fiber (PMSMF). Before the photon pair is coupled into the PMSMF, two dichroic mirrors and an interference filter separate the SPDC photons from the pump photons. The PMSMF guides the photon pair to a $4$-level rack [see Fig.~\ref{fig:setup} b)] made out of strut profiles that contains the detection module, a second computer for the source control, and a Raspberry Pi for GPS tracing.
The GPS antenna is placed at the window of the airplane. Fig.~\ref{fig:setup} a) shows the second level of the rack containing the detection module and the single-photon avalanche diodes.
In the detection module, the photons are spatially separated by a 50:50 cube beamsplitter. Post-selecting on those photon pairs which take different paths yields a Bell state $\ket{ \psi^{+}}=\tfrac{1}{\sqrt{2}} \bigl(  \ket{H}_{\mathrm{A}}\ket{V}_{\mathrm{B}} + \ket{V}_{\mathrm{A}}\ket{H}_{\mathrm{B}}      \bigr)$ directly after the beamsplitter,
where $H$ and $V$ denote the horizontal and vertical polarization, respectively, and the subscripts A and B indicate the spatial modes of the transmitted and reflected photons travelling towards Alice's and Bob's polarization analyzer, respectively.
A \mbox{quarter-,} half-, and quarter-wave plate arrangement (QHQ) in the reflected output enables tuning of the phase of the Bell state, which we set such that we obtain the anti-symmetric Bell state
\vspace*{-3mm}
\begin{align}
\ket{ \psi^{-}} = \frac{1}{\sqrt{2}} \left(  \ket{H}_{\mathrm{A}}\ket{V}_{\mathrm{B}} - \ket{V}_{\mathrm{A}}\ket{H}_{\mathrm{B}}      \right)\,.
\end{align}
Each polarization analyzer consists of a 50:50 beamsplitter with a half-wave plate (HWP) and a polarizing beamsplitter (PBS) in each output. A single-photon avalanche diode (SPAD) in each PBS output detects the photons. A time-tagging module (TTM) records the photon arrival times in each detector. Finally, detection events at the detectors of Alice and Bob recorded within a time window of 2~ns are identified as coincidences.
The crystal is pumped with a power of $22.2\pm0.1$~mW. Our detectors measure different single-photon count rates between $\sim 107$k and $\sim 167$k~counts/s.
The largest coincidence rate measured for any detector pair was around 6600 coincidences/s.
In addition, an accelerometer measures the acceleration along the $x$, $y$, and $z$ direction.
The optical components of the detection module are mounted in a rugged multicube system.
The opaque cube system prohibits stray light from entering the
photon-sensitive system, thus reducing inadvertent background counts. After alignment all optical components are additionally fixed with locking screws in order to prevent misalignment through forces caused by high accelerations or vibrations. The cubes are mechanically interconnected via steel rods. The stiff cube assembly is fixed to a 7~mm thick aluminium plate, which is in turn placed upon a cork pad in order to dampen aircraft vibrations.
The source box and the rack with the detection module are attached to rails on the floor of the Airbus A310, which itself has been modified for parabolic flights. The aircraft serves as a laboratory for experiments in micro- and hypergravity with accelerations of up to 1.8~g, where the term hypergravity describes accelerations above 1~g and microgravity refers to the acceleration regime from $10^{-2}$~g to $10^{-6}$~g.
Our experiment took place during the 77th ESA parabolic flight campaign from October 18th to October 29th 2021, with three flights, one each on the 26th, 27th, and 28th of October.
\begin{figure*}[ht!]
\centering
\includegraphics[width=0.88\textwidth]{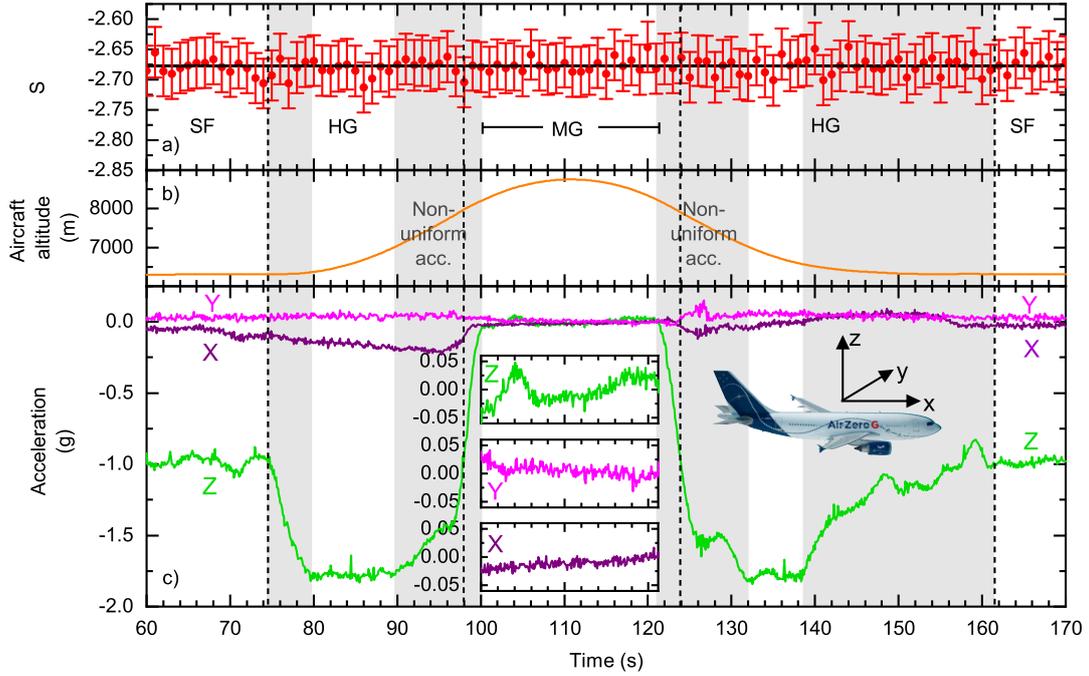}
\vspace*{-4mm}
\caption{\textbf{Bell-inequality violation during a typical parabolic flight manoeuvre.} \textbf{a)} the Bell-CHSH parameter $S$, \textbf{b)} aircraft altitude above sea level (in meters), and \textbf{c)} accelerations along the $x$ (dark magenta), $y$ (pink), and $z$ axes (green, all in units of g$=9.81~\mathrm{ms}^{-1}$) are shown as functions of the flight time (in seconds) for a 110~s interval during a typical parabolic flight manoeuvre. For better overview the 5 phases of the total time interval with (almost) uniform acceleration are shown with white background: steady flight (SF), hypergravity (HG), microgravity (MG), and again hypergravity (HG) and steady flight (SF). The regions with gray background correspond to the transitions between these 5 phases with non-uniform acceleration.
The Bell-CHSH parameter $S$ fluctuates around an average value of $-2.678$ during all phases of the manoeuvre. The integration time for each data point is 1~s.
For the coincidences (without subtraction of accidentals) we assume Poissonian statistics and we calculate the uncertainties $\Delta S$, the standard deviation of the mean of $S$, via Gaussian error propagation. Error bars correspond to three standard deviations of the mean. We note that the time stamps of acceleration values and S-values have a fixed but unknown relative shift of up to 1s.
During the hypergravity phases, the acceleration reaches values of up to 1.8~g.
The insets in c) show magnifications of the accelerations along the $x$ (green, top), $y$ (pink, middle), and $z$ axes (dark magenta, bottom) during microgravity. The aircraft image has been used with permission of Novespace.
}\label{fig:S_Acc_AccZoomIn_Altitude}
\end{figure*}
During a single flight the plane completes a series of 31 parabolic flight manoeuvres, each divided into three stages: at the beginning, the aircraft experiences a hypergravity phase (HG) reaching accelerations of up to 1.8~g along the $z$ direction. After about 22 seconds, the aircraft enters into a parabolic trajectory, while the aircraft experiences weightlessness. The transition period from hyper- to microgravity lasts around three seconds. However, due to air turbulence, the acceleration fluctuates around 0~g, see the inset in Fig.~\ref{fig:S_Acc_AccZoomIn_Altitude}~c). With microgravity periods of about 22~s in each parabola, the total time in microgravity during flight one amounts to $653 \, \mathrm{s}$ and during flight two to $687 \, \mathrm{s}$. Further details on the experimental setup and the flight can be found in Sec.~\ref{appendix:setup} of the appendix.\\[-2mm]


\noindent\textbf{Results}.\
During the flight we recorded the coincidence rates $C(\alpha,\beta)$ for all combinations of polarizer angles $\alpha = 0^\circ$ and $\alpha' = 45^\circ$ of Alice's detectors and $\beta = 22.5^\circ$ and $\beta' = 67.5^\circ$ of Bob's detectors to estimate the Bell-CHSH parameter
\begin{equation}
S = \left| E(\alpha, \beta) - E(\alpha, \beta') \right| + \left| E(\alpha', \beta) + E(\alpha', \beta') \right|,
\end{equation}
from the correlation functions $E(\alpha, \beta)$ given by
\begin{equation}
E(\alpha, \beta) = \frac{C(\alpha, \beta)+C(\overline{\alpha}, \overline{\beta})-C(\alpha, \overline{\beta})-C(\overline{\alpha}, \beta)}{C(\alpha, \beta)+C(\overline{\alpha}, \overline{\beta})+C(\alpha, \overline{\beta})+C(\overline{\alpha}, \beta)},
\end{equation}
where the overline indicates an orthogonal polarizer direction, e.g., $\overline{\alpha}\perp\alpha$.
For any local realistic theory, the value of $S$ is bounded by $|S|\leq 2$,
while quantum mechanics predicts values up to a maximum of $2\sqrt{2}$.
Our setup with two analyzer modules each for Alice and Bob permits us to measure all four correlation functions
without intermittent switching between different angles.\\[-4mm]

On the first flight, time-tagged data was collected
over the course of 30 parabolic flight manoeuvres.
Figure~\ref{fig:S_Acc_AccZoomIn_Altitude} a) shows the resulting Bell-CHSH parameter $S$ for a 110~s time window
covering one entire parabola (the first parabola for which data was recorded during the flight on 26th October 2021), featuring
accelerations of up to 1.84~g and a microgravity phase lasting 21~s.
For each data point $S_{t}$, the coincidence measurements were integrated over 1~s.
The coincidences were assumed to have a Poissonian distribution and the standard deviations for $S$ were calculated using error propagation.
The aircraft altitude and acceleration (in the $x$, $y$, and $z$ directions) are shown in Figs.~\ref{fig:S_Acc_AccZoomIn_Altitude} b) and c), respectively.
The average Bell-CHSH parameter $\overline{S}\suptiny{0.5}{-2}{(1)}= \tfrac{1}{111}\sum_{t=60 \mathrm{s}}^{t=170\mathrm{s}}~S_{t}$ (horizontal black line) across the entire parabola (from $t = 60 \, \mathrm{s}$ to $170 \, \mathrm{s}$, with associated average standard deviation of the mean $\overline{\Delta S}\suptiny{0.5}{-2}{(1)}= \tfrac{1}{111}\sum_{t=60 \mathrm{s}}^{170\mathrm{s}} \Delta S_{t}$) is $-2.678\pm 0.013$, corresponding to an average distance of $\overline{\sigma}\suptiny{0.5}{-1}{(1)} = \tfrac{1}{111} \sum_{t = 60 \mathrm{s}}^{170 \mathrm{s}} |S_{t}-(-2)|/\Delta S_{t}=48.9$ standard deviations from the local-realistic bound of~-2.\\[-3mm]

\begin{figure*}[ht!]
\centering
\includegraphics[width=0.48\textwidth]{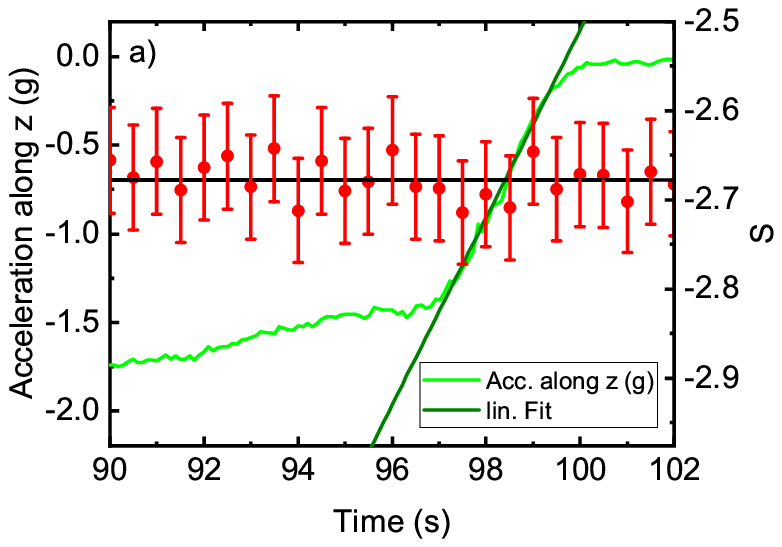}
\includegraphics[width=0.48\textwidth]{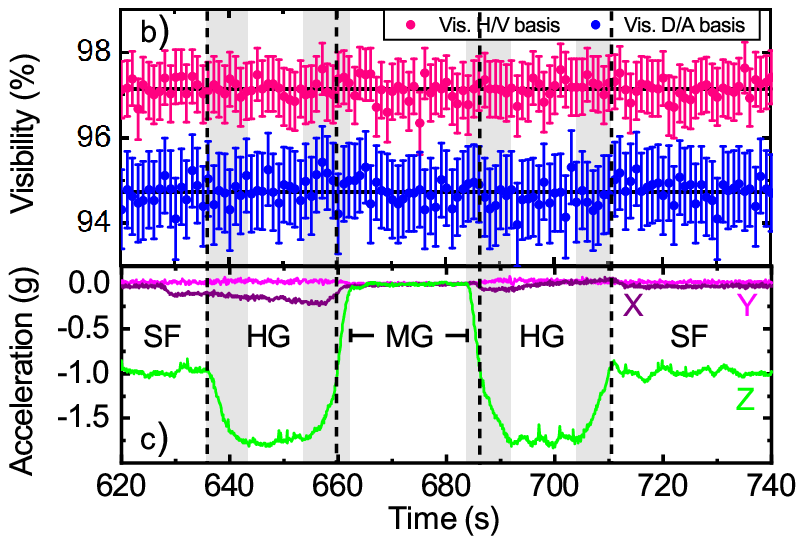}
\vspace*{-6mm}
\caption{\textbf{a)\ Bell-inequality violation during the transition from hyper- to microgravity.}
The Bell-CHSH violation (red, vertical axis on the right-hand side) and acceleration along the $z$ direction (green, vertical axis on the left-hand side, in units of g) are shown as functions of time (in seconds). The data is that used for the generation of Fig.~\ref{fig:S_Acc_AccZoomIn_Altitude}, but reduced to the time interval from $t = 90$~s to $t = 102$~s and a shorter integration time ($0.5$~s instead of 1~s) for the calculation of $S$.
The black horizontal line shows the time-averaged value $S = -2.678$ for the interval from $t = 60$~s to $t = 170$~s. The error bars correspond to three standard deviations of the mean. The straight dark green line shows a linear fit to the acceleration data in the even narrower time window from $t=97.02$ s to $t=99.34$ s.
The slope of the linear fit
indicates a jerk (time-derivative of the acceleration) in the $z$ direction of $j = 0.5305 \pm 0.0234$ $\mathrm{g}/\mathrm{s}$ ($\mathrel{\hat=} 5.204 \pm 0.229$ $\mathrm{m}/\mathrm{s}^3$).
\textbf{b)\ Visibilities in the H/V and D/A bases during a parabolic flight manoeuvre}.
The visibilities in the H/V ($0^\circ/90^\circ$, magenta, top) and D/A ($45^\circ/135^\circ$, blue, bottom) bases, measured during another parabola, fluctuate around average 
values of $V_{\mathrm{HV}} = 97.15 \%$ and $V_{\mathrm{DA}} = 94.73 \%$ (time-averaged from $t = 620$ s to $t = 740$ s) respectively during different accelerations in the range between 1.8 g and microgravity.
\textbf{c)\ Accelerations} along the $x$ (dark magenta), $y$ (pink), and $z$ (green) directions are shown for the selected time window.
The intervals with gray background indicate the phases of significant non-uniform acceleration as in Fig.~\ref{fig:S_Acc_AccZoomIn_Altitude}.}
\label{fig:Plot_S_acc_Jerk}
\end{figure*}

The collected data shows that the CHSH inequality is (strongly) violated and entanglement thus clearly preserved during all periods of uniform and non-uniform acceleration, with accelerations between -1.84~g and 0~g. We observe this not just for the parabola for which data is shown in Fig.~\ref{fig:S_Acc_AccZoomIn_Altitude}, but also for all 29 subsequent parabolas: $S$ only shows small fluctuations around the average value of -2.680 over the whole flight,
with a maximum value $S_{\mathrm{max}}$ = $-2.6202$
and a minimum value $S_{\mathrm{min}}$ = $-2.7323$ (see Fig.~\ref{fig:All_S_values_over_complete_flight} in the appendix).
The average standard deviation of the means across all $S$ values $S_{t}$ is
$\overline{\Delta S} = \tfrac{1}{6843} \sum_{t} \Delta S_{t} = 0.0140$. We note that with decreasing integration time, the fluctuations of the CHSH-parameter $S$, and the standard deviation become bigger. Based on the data collected during the flight, we estimate a limit for the integration time of $t_{\mathrm{int}} < 3.9$~ms below which the amplitude of the error bars with 3 standard deviations becomes bigger than the difference $|\overline{S}-(-2)|$.\\[-2mm]

In our data we do not observe any statistically relevant influence of the different levels of acceleration on the Bell-inequality violation, including the non-uniformly accelerated transitions from hyper- to microgravity and back. An example of such a transition with the corresponding Bell-CHSH violations is shown in Fig.~\ref{fig:Plot_S_acc_Jerk}~a). There, a linear fit provides an estimate for the jerk, the change of acceleration in time, of
$j=0.5305 \pm 0.0234~\mathrm{g}/\mathrm{s}$ ($\mathrel{\hat=} 5.204 \pm 0.229~\mathrm{m}/\mathrm{s}^3$). The value of $S$, here with an integration time of 500 ms, fluctuates around the time-averaged value of $S = -2.678$ without a significant change due to the jerk compared to the values of $S$ during steady flight (with a gravitational pull similar to that on the ground).\\[-3mm]

In order to statistically underpin the observation that the different levels and changes of acceleration have no influence on the polarization entanglement we compare our $S$-value data from pairs of flight segments with different levels of accelerations using the two-sample Kolmogoroff-Smirnoff test (KST), which tests the hypothesis that the two samples follow the same underlying probability distribution (see Appendix~\ref{appendix:KST} for more details). Using these tests we find no experimental indication (with a 5~\% significance level) for an influence of the acceleration on the polarization entanglement in our setup. In all performed KSTs the null hypotheses, claiming that acceleration and a change of acceleration have no influence on polarization entanglement are confirmed except for one case. In this specific case, we ascribe the fact that the maximum distance between the empirical cumulative distribution functions of $S$ values measured during microgravity and $S$ values measured during a phase of changing accelerations is bigger than the critical value, to a small sample size of $11$.\\[-3mm]

During a further sequence of 31 parabolic flight manoeuvres on 27th October 2021, the visibilities in the H/V and D/A bases corresponding to horizontal/vertical and diagonal/anti-diagonal polarizer settings, $0^\circ/90^\circ$ and $45^\circ/135^\circ$, respectively, were measured
an additional figure of merit. These visibilities can be used to detect and in principle even quantify entanglement (via lower bounds to the entanglement of formation,  see, e.g., \cite{BavarescoEtAl2018,FriisVitaglianoMalikHuber2019}).
The visibilities can be obtained from the respective coincidences of the detectors of Alice and Bob via
\begin{equation}
V_{\mathrm{H/V}} = \frac{ C_{\mathrm{HV}} + C_{\mathrm{VH}} - C_{\mathrm{HH}} - C_{\mathrm{VV}} }{ C_{\mathrm{HV}} + C_{\mathrm{VH}} + C_{\mathrm{HH}} + C_{\mathrm{VV}} },
\end{equation}
and in complete analogy $V_{\mathrm{D/A}}$ is obtained from $C_{\mathrm{DA}}$, $C_{\mathrm{AD}}$, $C_{\mathrm{DD}}$, and $C_{\mathrm{AA}}$.
Fig.~\ref{fig:Plot_S_acc_Jerk} b) shows the visibilities in both bases during a single parabola. Each data point is the result of integration over 1~s. For the coincidences we again assume Poissonian statistics and calculate error bars using Gaussian error propagation.
The time-averaged visibilities (from $t = 620 \, \mathrm{s}$ to $t = 740 \, \mathrm{s}$) are $V_{\mathrm{HV}} =97.15$~\% and $V_{\mathrm{DA}} =94.73$~\% in the H/V and D/A basis, respectively.
These results show in more detail that our setup for entanglement generation and detection is insensitive against aircraft vibrations (see also Fig.~\ref{fig:All_Visibilities_values_over_complete_flight}~and~\ref{fig:All_Visibilities_over_complete_flight_6_rows} in the appendix) and accelerations with peak values of up to 1.99 g (here our backup accelerometer measured $2.090 \pm 0.002 \, \mathrm{g}$ while the accelerometer of the aircraft measured 2.07~g, see Fig.~\ref{fig:Comp_three_Accelerometers} in the appendix), at which the acceleration sensor saturates. Our setup provides strongly polarization-entangled photon pairs in this tough ambience over 30 parabolas for 1.9 hours.\\[-4mm]

\noindent\textbf{Effects of non-uniform motion and their tests}.\ To provide some context for our experiment and its results, we will now briefly discuss potential quantum effects connected to strong accelerations and their tests. A fundamental effect related to non-uniform is the dynamical Casimir effect (for a recent review see, e.g, \cite{Dodonov2020}) and related parametric amplification effects caused by non-stationary boundary conditions for quantum fields~\cite{BruschiFuentesLouko2012,FriisLeeLouko2013,AlsingFuentes2012,FriisPhD2013}. Such effects are usually modelled as Gaussian transformations of the field modes corresponding to squeezing (pair creation to lowest order in the relevant parameters) and shifts of excitations. However, the required proper accelerations, or temporal changes thereof, to make the predicted effects visible in experiments are many orders of magnitude away from what can practically be achieved with mechanical motion. Yet, the modulation of microwave-field boundary conditions represented by superconducting quantum interference devices (SQUIDs) can be carried out much more rapidly, which led to the landmark demonstration of the dynamical Casimir effect in~\cite{WilsonJohanssonPourkabirianJohanssonDutyNoriDelsing2011}.\\[-3.5mm]

In the experiment we present here, the (changes of) accelerations are far too small to expect any effects akin to the dynamical Casimir effect. Nevertheless, our setup provides a testbed for demonstrating the robustness of entanglement-detection experiments under extreme but still reasonably well-controlled conditions that allow ruling out any influence that the non-uniform motion might have.\\[-3.5mm]

Previous work in this direction~\cite{Fink} has reported measurements of the visibility of polarization-entangled photon pairs for different levels of uniform acceleration, ranging from microgravity during free fall in a drop tower of 12~m height to uniform accelerations of up to 30~g generated by a centrifuge. The integration time in the drop tower and thus the time in which the photons experience microgravity is only 1.56 s. For the purpose of a higher accuracy, a longer integration time would be necessary, which could be achieved in two ways in a drop tower: increased tower height or a sufficient number of repetitions. Both options entail a high risk of damaging components of the setup, in particular the optical elements and electronic devices.\\[-3.5mm]

Meanwhile, centrifuges allow one to create strong centripetal forces but do not permit tests of the setup during microgravity due to the gravitational acceleration in the direction parallel to the rotation axis.
Nevertheless, setups based on rotational motion have been suggested as platforms for revealing and concealing entanglement by rotational motion \cite{Conc_Rev_Etglm}, and have been used to investigate the dependence of the indistinguishability of single photons that gives rise to Hong-Ou-Mandel interference on the angular velocity of the rotating optical system \cite{HOM_Rot}. Our experiment goes beyond previous approaches using drop towers and centrifuges in that it allows for the uninterrupted observation of the Bell-inequality violation over long integration times during which the acceleration continuously changes but is at any given time the same for all components of the experiment.\\[-3.5mm]

Alternative platforms for microgravity experiments are space stations like the ISS or satellites.
To the best of our knowledge no experiments of the type discussed here have been conducted aboard the ISS, but sources of entangled photon pairs have been installed on the satellite Micius~\cite{MiciusI,MiciusII} and on a CubeSat nano-satellite~\cite{VillarEtAl2020}, both in low Earth orbit. By using the source aboard Micius to distribute photons to Earth via telescopes it has been demonstrated that entanglement persists for photons spatially separated by 1200 km. While the photons are in this case generated in microgravity, the detection takes place on Earth in 1~g \cite{MiciusI}, in contrast to our setup, where the acceleration is the same for all components of the experiment but varies with time. A further experiment with the entangled-photon source on Micius describes the distribution of two photons where one photon is analyzed on the satellite and the other one is sent to a receiving station on Earth \cite{MiciusII}. For the CubeSat source~\cite{VillarEtAl2020} the violation of a Bell inequality was reported, but no information was provided regarding the satellite`s motion.\\[-3.5mm]

\noindent\textbf{Summary and conclusion}.\
In summary, we built a compact setup for the generation and detection of polarization-entangled photon pairs and installed it into a modified Airbus A310. During a sequence of parabolic flight manoeuvres of the 77th ESA parabolic flight campaign, we demonstrated that our optical setup is sufficiently robust against aircraft vibrations and accelerations (between 0 and 1.99~g) to allow for the continuous strong violation of a Bell inequality, indicating that the strength of polarization entanglement persists during these accelerations. Neither the hypergravity phases with peak acceleration values of up to approximately 1.99~g, nor the microgravity phases (lasting around 20-22~s) had a statistically relevant influence on the quantity $S$. In a next sequence, the visibilities of the polarization entangled photons in the H/V and D/A bases were measured, which also showed no dependence on the motion of the aircraft, persisting in the phases of micro- and hypergravity, and during the transitions between them.\\[-3mm]

Our results provide further evidence for the feasibility of entanglement-based quantum communication and its applications outside temperature stabilized and vibration-protected laboratories. In addition, our experiment adds another important reference point for tests of entanglement and quantum-information processing under the influence of non-inertial and non-uniformly accelerated motion for acceleration levels within the tolerance limits of human experimenters.
For future investigations of possible influences of stronger (non-uniform) acceleration on entanglement, we envisage experiments performed in centrifuges whose angular velocities can be continuously adjusted in a larger range.


\bibliographystyle{apsrev4-1fixed_with_article_titles_full_names_new}
\bibliography{references}


\newpage
\noindent\textbf{Acknowledgments}.\
We thank Roland Blach for building the components for the rack and preparing the Zarges box for the experiment. We thank Daniel Hinterramskogler who accompanied us with his camera and took
pictures and videos during the installation of the experiment in the aircraft.
We want to thank Nicolas Courtioux, Thomas Villatte, Frédéric Gai, and the whole team of Novespace, who accompanied us for six months during the preparation with their support to make this experiment
possible.
Further, we want to thank Neil Melville, Nigel Savage, and the European Space Agency who made this
experiment possible.
We acknowledge the Austrian Academy of Sciences in cooperation with the FhG ICON-Program "Integrated Photonic Solutions for Quantum Technologies (InteQuant)".
N.F. acknowledges support from the Austrian Science Fund (FWF) through the project P~31339-N27 and through the project P 36478-N funded by the European Union - NextGenerationEU, as well as from the Austrian Federal Ministry of Education, Science and Research via the Austrian Research Promotion Agency (FFG) through the flagship project HPQC (FO999897481) funded by the European Union – NextGenerationEU.
M.H. acknowledges funding
from the European Commission (grant
’Hyperspace’ 101070168) and from the European Research Council (Consolidator grant ’Cocoquest’ 101043705).\\[2mm]

\noindent\textbf{Author contributions}.\
J.B. planned and organized the flight, built the setup, and analyzed the data.
L.B. contributed to the data-analysis and helped to build the setup.
M.F. built the original photon-pair source which was modified by J.B.
J.B., L.B., S.E., S.N. tested the setup in advance in Vienna.
J.B., L.B., S.E., S.N., M.B., installed the setup into the aircraft and performed the experiment during the parabolic flights.
J.B., M.H. and N.F. wrote the manuscript. All authors contributed to discussions of the results and the manuscript.
R.U. conceived and initiated the experiment.\\[2mm]

\newpage\noindent\textbf{Rights Retention Statement}.\ This research was funded in whole or in part by the Austrian Science Fund (FWF) [\href{https://doi.org/10.55776/P36478}{10.55776/P36478}]. For open access purposes, the author has applied a CC BY public copyright license to any author accepted manuscript version arising from this submission.



\newpage
\clearpage
\begin{widetext}

\hypertarget{sec:appendix}
\appendix

\section*{Appendix: Supplemental Information}

\renewcommand{\thesubsubsection}{A.\Roman{subsection}.\arabic{subsubsection}}
\renewcommand{\thesubsection}{A.\Roman{subsection}}
\renewcommand{\thesection}{}
\setcounter{equation}{0}
\numberwithin{equation}{section}
\setcounter{figure}{0}
\renewcommand{\theequation}{A.\arabic{equation}}
\renewcommand{\thefigure}{A.\arabic{figure}}


\noindent In this appendix we present additional details for our experiment and the data analysis. The appendix is structured as follows: in Sec.~\ref{appendix:setup} we present additional information about the setup and the zero-g flight as well excerpts from the single-photon count rates and coincidence rates. In Sec.~\ref{appendix:KST} we provide a detailed description of the Kolmogoroff-Smirnoff test (KST) that we use to test the hypothesis that the accelerations have no effect on the Bell-inequality violation. Finally, Sec.~\ref{appendix:parabola data} shows the complete results for all $S$-value measurements and visibility measurements over all 30 and 31 parabolas for the first and second flight days, respectively.


\vspace*{-3mm}
\subsection{Setup and Flight Details}\label{appendix:setup}
\vspace*{-2mm}

\noindent Figure~\ref{fig:setup_in_aircraft} shows a picture of our setup installed into the aircraft cabin. The setup features two computers. One for the source control (PC1) and one for data acquisition (PC2). PC1 additionally records the temperature of the non-linear crystal and the laser current, and is equipped with an accelerometer. PC2 records the time stamps of the single-photon detection, the acceleration data of a second accelerometer, the cabin pressure, and the temperature in the
3rd level of the rack (see Fig.~\ref{fig:setup} of the main text). The routes of both flights are depicted in the maps shown in Fig.~\ref{fig:Maps}.\\[-3mm]

\begin{figure}[h!]
\centering
\includegraphics[width=0.42\textwidth]{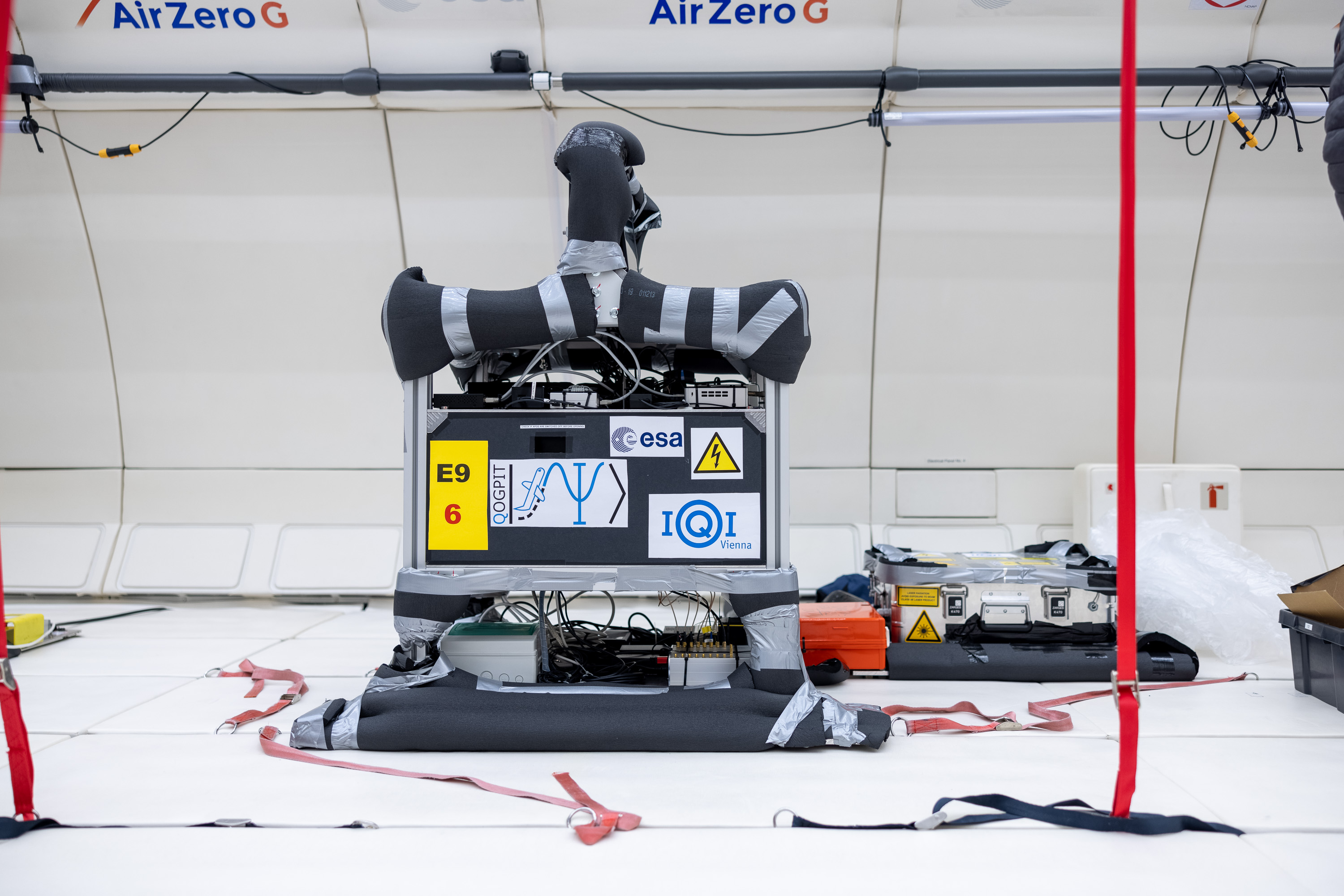}
\caption{
\textbf{Setup installed in the aircraft}. In the centre of the image the rack (745 mm $\times$ 545 mm $\times$ 1230 mm) containing the components for the measurements can be seen. Black panels cover the 2nd level of the rack which contains the detection modules as shown in Fig.~\ref{fig:setup} of the main text. On the right-hand side of the rack the aluminium box (600 mm $\times$ 400 mm $\times$ 250 mm) containing the source for photon-pair generation can be seen fixed to the floor of the aircraft cabin. The red straps in the image are used to secure the experimenters during the microgravity phases. Picture taken by Daniel Hinterramskogler.
}
\label{fig:setup_in_aircraft}
\end{figure}
\vspace*{-4mm}
\begin{figure*}[hb!]
\centering
\includegraphics[width=0.42\textwidth]{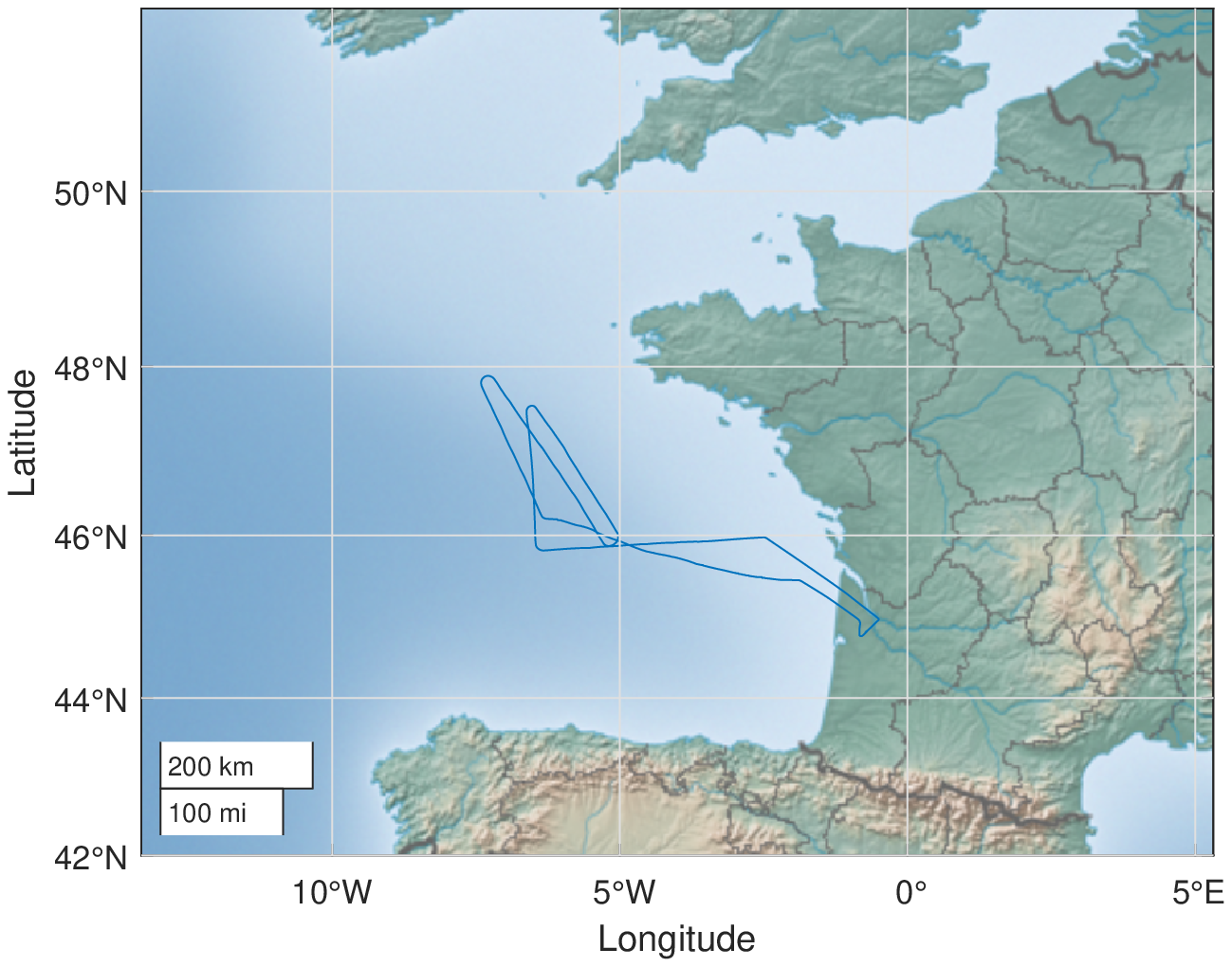}
\includegraphics[width=0.42\textwidth]{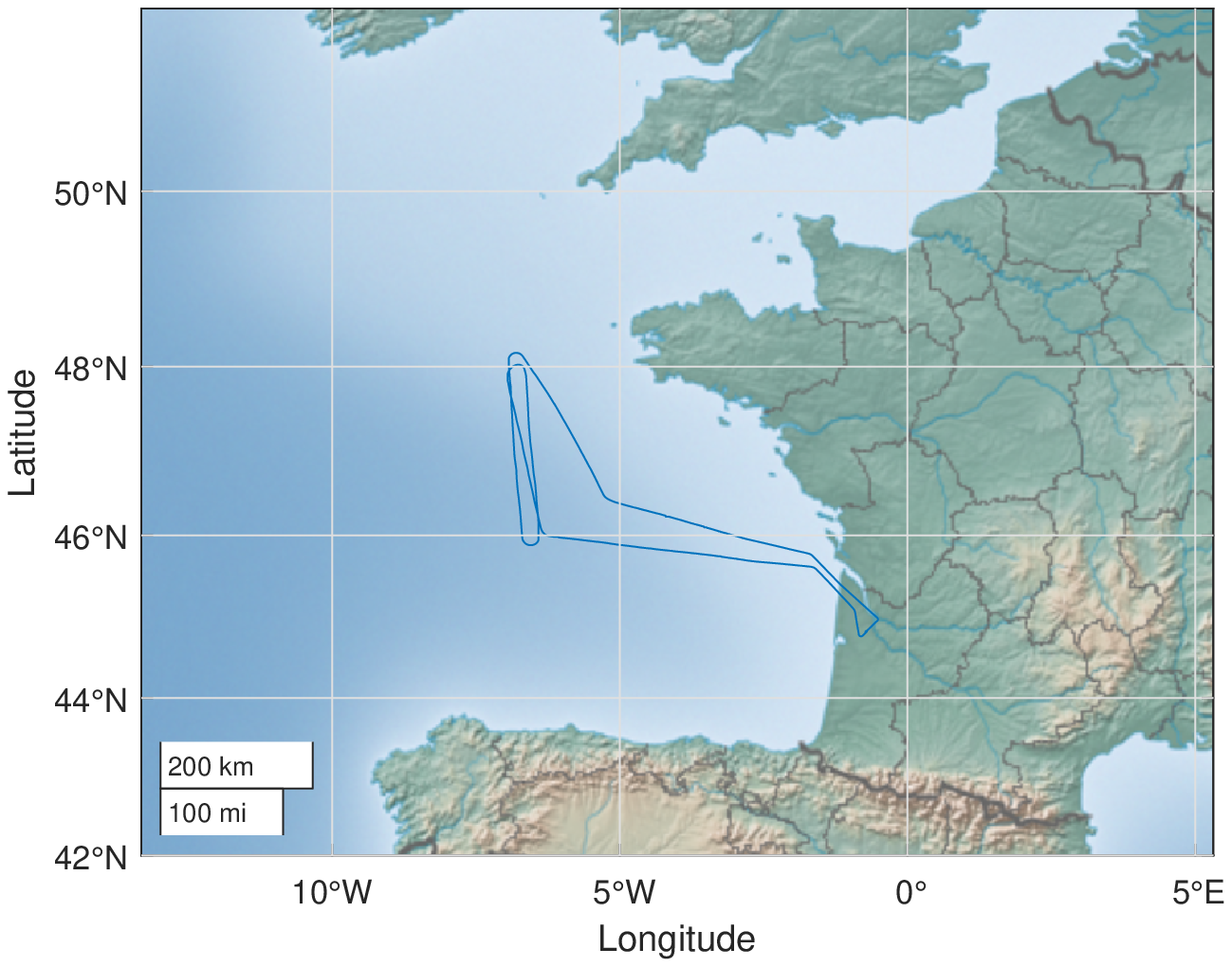}
\vspace*{-2.5mm}
\caption{\textbf{Flight trajectories}.
The images show maps (Mercator projection) of the west coast of France and the Atlantic Ocean (Bay of Biscay) with the flight trajectory of the first flight on 26th of October shown on the left-hand side and that of the second flight on 27th of October on the right-hand side.
The flights started from the airport in Bordeaux Merignac and ended there.
The parabolic flight manoeuvres started and ended at an altitude of $6300 \pm 100$~m. The measured apex of the parabolas was between 8684 m and 8961 m above sea level. This data was provided by Novespace.
}
\label{fig:Maps}
\end{figure*}

\clearpage


\vspace*{-4mm}
\subsubsection{Acceleration measurements}
\vspace*{-2mm}

\noindent Both of our accelerometers are placed directly next to the detection module in the 2nd level of the rack. Additionally, a third accelerometer installed in the aircraft itself provides us with additional acceleration data. On the one hand, the second accelerometer acts as a backup device, and on the other hand, its data is used to synchronize the measured data from PC1 with the measured data of PC2.
Figure~\ref{fig:Comp_three_Accelerometers} shows the absolute values of the accelerations recorded by the three accelerometers during the first parabola on the first flight.\\[-3mm]

In the following and in the main text we use the acceleration data of the accelerometer that is connected with PC2. More details on the values of the accelerations during selected individual parabolas can be seen in Fig.~\ref{fig:S_Acc_AccZoomIn_Altitude} and Fig.~\ref{fig:Plot_S_acc_Jerk} of the main text, while the data for the vertical accelerations over all parabolas of both days can be found in Sec.~\ref{appendix:parabola data}.\\[-3mm]

Each of the parabolic flight maneuvers is then divided into 3 different phases according to the measured accelerations: an initial hypergravity phase, then a microgravity phase, and again a phase of hypergravity. In between these main phases there are transition phases with significant changes in acceleration. During the initial hypergravity phases of the parabolic flight maneuvers, the acceleration reaches values of 1.8 g and even up to 1.99 g (here our backup accelerometer measured $2.090 \pm 0.002 \, \mathrm{g}$ while the accelerometer of the aircraft measured 2.07~g, parabola 4, flight one), with our sensor saturating at 1.99 g (accelerometer connected to PC2).
The microgravity phases last between 20.4~s and 23.7~s. For the first and second flight day we recorded average durations of $\overline{t}_{\mathrm{microgravity}} =21.78 \pm 0.12$~s and $\overline{t}_{\mathrm{microgravity}} = 22.15 \pm 0.15$~s, respectively, for the microgravity phases of the individual parabolas. The total times spent in microgravity (in which we effectively measured) on the first and second flight were recorded to be 653~s and 687~s, respectively.
In comparison, the average times spent in hypergravity during any of the parabolas of the first and second flight were $\overline{t} = 51.06 \pm 1.71$~s and $\overline{t} = 49.50 \pm 1.62$~s, respectively. \\[-3mm]

The accelerometer of PC2 records the data of acceleration on average every 0.1049 s. The time difference between each recorded value fluctuates. The minimum time difference is $\triangle t_{\mathrm{min}}=0.0999$~s and the maximum time difference is $\triangle t_{\mathrm{max}}=0.2040$~s. The maximum jerk measurements are limited by the minimum time difference of two consecutive acceleration data points and the maximum acceleration values. Acceleration values are recorded in a range between -2~g and 2~g with a 14-bit resolution of the sensor. With this data we calculate a maximum jerk that we can measure theoretically:
\begin{equation}
    j_{\mathrm{max,possible}} = \frac{\triangle a_{\mathrm{max}}}{\triangle t_{\mathrm{min}}} = \frac{4 \mathrm{g}}{0.0999\mathrm{s}} = 40.04\frac{\mathrm{g}}{\mathrm{s}}\,.
\end{equation}
\\[-3mm]

The second accelerometer, connected to PC1, is made by a different manufacturer. It records the data of acceleration on average every 0.1049 s. The minimum and maximum time difference are $t_{\mathrm{min}}=0.0070$~s and $t_{\mathrm{max}}=0.5610$~s, respectively. Like the other acceleration sensor, with this sensor we also measured in a range of $\pm 2$~g. The sensor has a resolution of 14 bit. The maximum jerk is:
\begin{equation}
    j_{\mathrm{max,possible}} = \frac{\triangle a_{\mathrm{max}}}{\triangle t_{\mathrm{min}}} = \frac{4 \mathrm{g}}{0.0070\mathrm{s}} = 571.43\frac{\mathrm{g}}{\mathrm{s}}\,.
\end{equation}
\\[-3mm]

\begin{figure*}[h!]
\begin{center}
\hspace*{2mm}\includegraphics[width=1.05\textwidth]{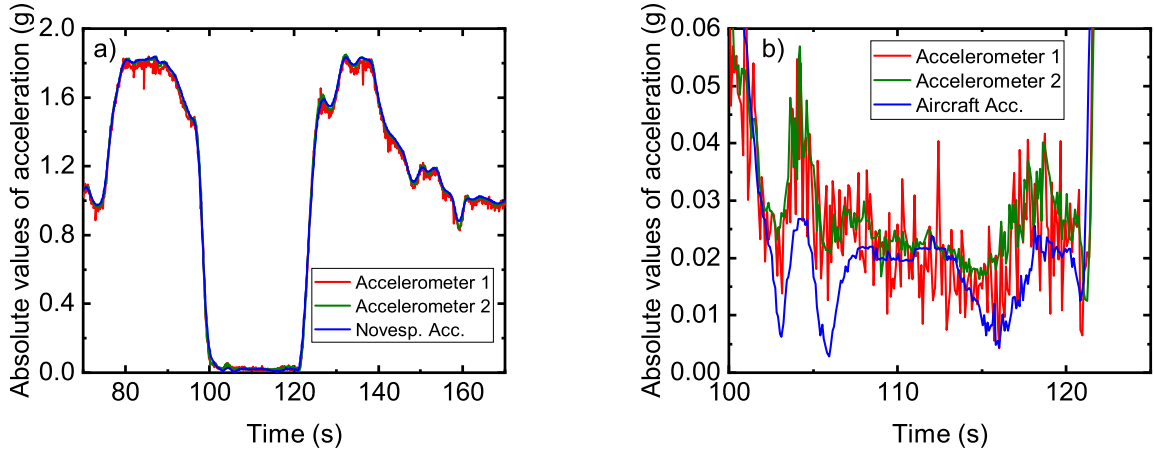}
\end{center}
\vspace*{-7mm}
\caption{\textbf{Acceleration during first parabola}. The graphs show the absolute values of the accelerations measured by the three different accelerometers during our experiment. Accelerometers 1 (red line) and 2 (backup accelerometer, green line) are placed within the 2nd level of the rack next to the detection module. The blue line shows the measured acceleration of the aircraft accelerometer (data from Novespace). In \textbf{a)} one can see the acceleration of the first parabola on day one. \textbf{b)} shows a magnification of the accelerations in \textbf{a)} for the microgravity phase from $t = 100 \, \mathrm{s}$ to $t = 125 \, \mathrm{s}$. In the main text we used the data of accelerometer 1. Accelerometer 2 had an offset of $-0.0559 \, \mathrm{g}$ along the y-axis which has been corrected in the data shown in this figure.
}
\label{fig:Comp_three_Accelerometers}
\end{figure*}


\vspace*{-6mm}
\subsubsection{Pressure and temperature during the parabolas}
\vspace*{-2mm}

\noindent During the flights, we measured the temperature of the non-linear crystal in the source, as well as the air pressure and air temperature in the aircraft cabin, see Fig.~\ref{fig:Cryst_Temp_Pressure_Cabin_Temp}. During both flights, the crystal temperature remained essentially constant, while the air temperature increased over time. During the first and second flight the temperature sensor one (two) measured a difference between the maximum and minimum air temperature of $\thicksim4.00^\circ \mathrm{C}$ $(4.54^\circ \mathrm{C})$ on day one and $\thicksim6.83^\circ \mathrm{C}$ $ (7.79^\circ \mathrm{C})$ on day two, respectively.

\begin{figure*}[ht!]
\centering
\hspace*{5mm}\includegraphics[width=0.98\textwidth]{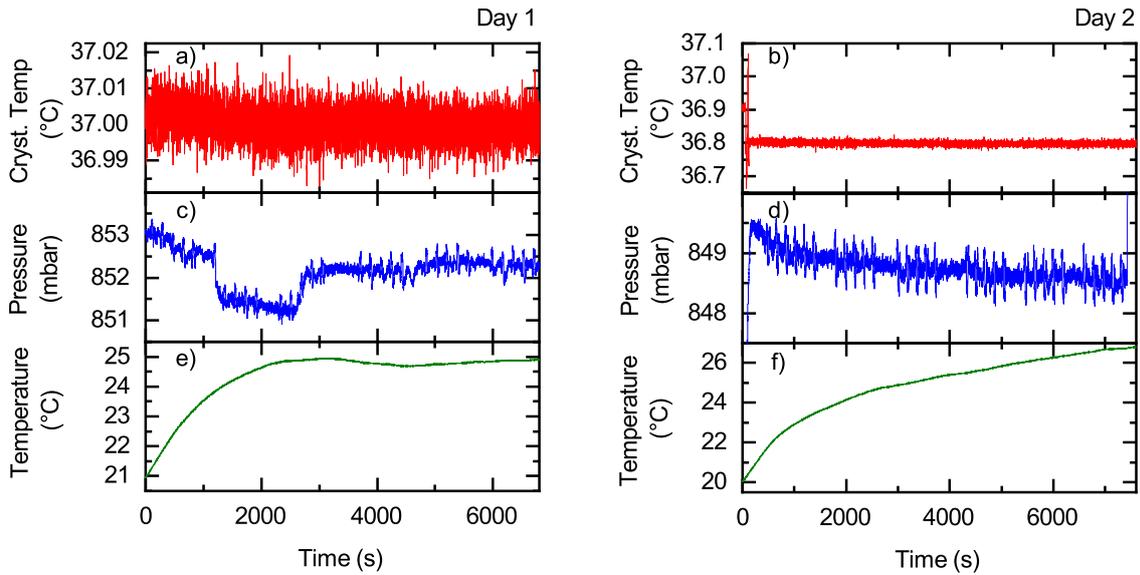}
\vspace*{-5mm}
\caption{\textbf{Temperature and pressure}.
\textbf{a)} and \textbf{b)} show the temperatures of the ppKTP crystal on the first and second day, respectively. Before the parabolic flight maneuvers started, from $t = 50 \, \mathrm{s} $ to $t = 127  \, \mathrm{s} $ we adjusted the crystal temperature, which only showed small fluctuations on both days. \textbf{c)} and \textbf{d)} show the pressures measured in the 3rd level of the rack. \textbf{e)} and \textbf{f)} show the temperatures measured in the 3rd level of the rack, which increased over time on both days.
}
\label{fig:Cryst_Temp_Pressure_Cabin_Temp}
\end{figure*}

\subsubsection{Optomechanics used for optic experiments in tough environments}
In the given dynamical parameter range the optomechanics show no detrimental effects on the measurements of the CHSH-Bell parameter $S$ and of the visibilities. Merely the single-photon count rates in Fig.~\ref{fig:Singles_Acc_over_time} and the coincidence rates in Fig.~\ref{fig:Coinc_over_time} show changes, which correlate with the change of acceleration, but do not seem to affect the $S$-value measurement nor the measurement of the visibility. We attribute this fact to the usage of mounts with additional locking screws. Thus, a misalignment of optical components attached to springs or ball-bearings is prohibited. It would be interesting to know up to which accelerations and jerks the optomechanics show the same performance as in our experiment. For this purpose, we suggest further experiments in centrifuges with accelerations above 30~g and with different jerks.\\

\subsubsection{Single-photon count rates and coincidence count rates during different levels of acceleration}
\vspace*{-2mm}

\noindent Although the $S$ values and, in particular, the visibilities in the H/V and D/A bases are not constant in time, they show no dependence on the acceleration (see Appendix~\ref{appendix:parabola data}).
However, for some of the detectors, we observe correlations between the single-photon count rates and the accelerations, as well as resulting correlations between the coincidence count rates and the accelerations. In Fig.~\ref{fig:Singles_Acc_over_time} these are showcased using plots of 3-second moving averages of the single-photon count rates. The count rates of detectors 4, 6 and 7 are compared with the acceleration as functions of time in Fig.~\ref{fig:Singles_Acc_over_time}~b), from which
it is apparent that, in particular, detector 6 shows fluctuations that coincide with periods of significant changes in the acceleration.
However, this behaviour is not shared by all detectors. For instance, the single-photon count rates of detector 4, also shown in Fig.~\ref{fig:Singles_Acc_over_time}~b) for comparison, seem to be largely unaffected by the acceleration.
We attribute these changes in the single-photon count rates to mechanical effects in the detection module.
We also observe similar fluctuations correlated with changes in the acceleration for some of the coincidence rates, see Fig.~\ref{fig:Coinc_over_time}, but the $S$ values derived from these coincidences are not affected by this effect.

\begin{figure*}[h!]
\centering
\hspace*{5mm}
\includegraphics[width=0.45\textwidth]{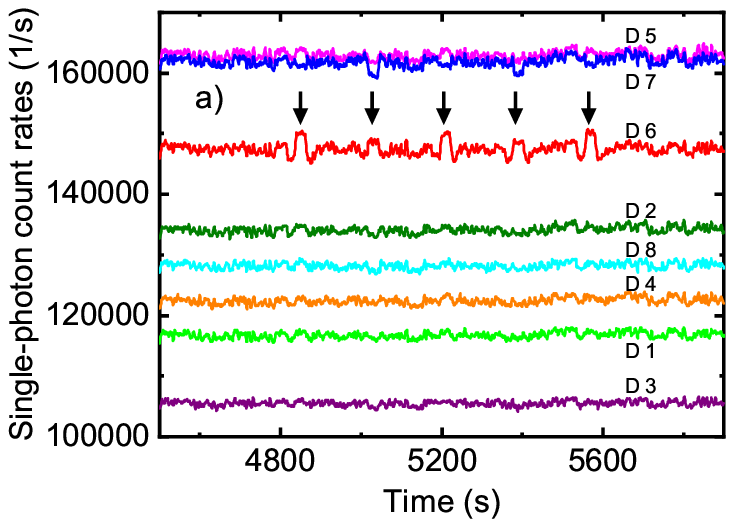}
\includegraphics[width=0.45\textwidth]{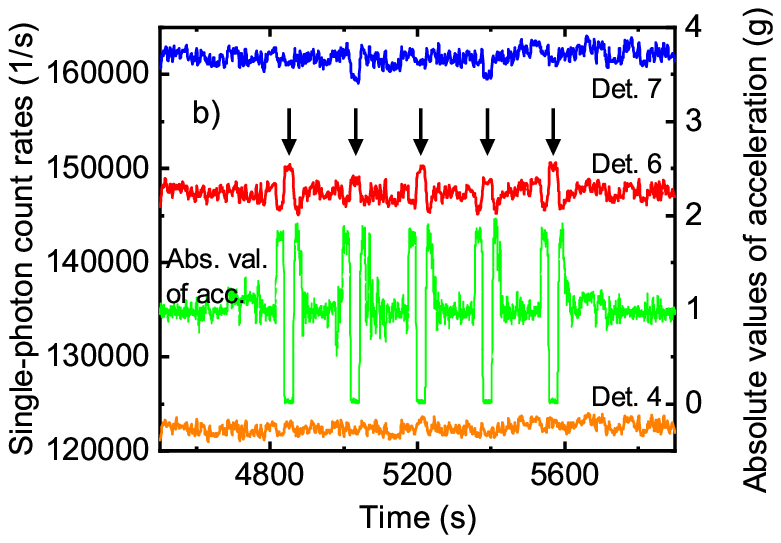}
\vspace*{-4mm}
\caption{
\textbf{Comparison of single-photon count rates and accelerations}.
\textbf{a)} The moving averages (with 3~s sliding window) of single-photon count rates of detectors D1 to D8 are shown for the time interval corresponding to parabolas 21 to 25 (first flight). \textbf{b)} The moving averages of the single-photon count rates (vertical axis on left-hand side) of detectors D7 (blue, top), D6 (red, second from the top) and D4 (orange, bottom), are compared with the acceleration (green, second from the bottom, vertical axis on right-hand side) measured at the same time. Black arrows indicate time intervals at which the count rates of D6 (red, second from the top) show qualitative correlations with low accelerations (microgravity). This effect is also visible, albeit more weakly, for D7 (blue, top). For other detectors (D4 shown as an example) this effect was not observed.
We attribute these correlations to a mechanical effect within the detection module.
Interestingly, while coincidences of certain detector pairs also show these correlations, see Fig.~\ref{fig:Coinc_over_time}, the $S$ values and visibilities do not.
}
\label{fig:Singles_Acc_over_time}
\end{figure*}




\begin{figure*}[h!]
\centering
\hspace*{3mm}
\includegraphics[width=0.48\textwidth]{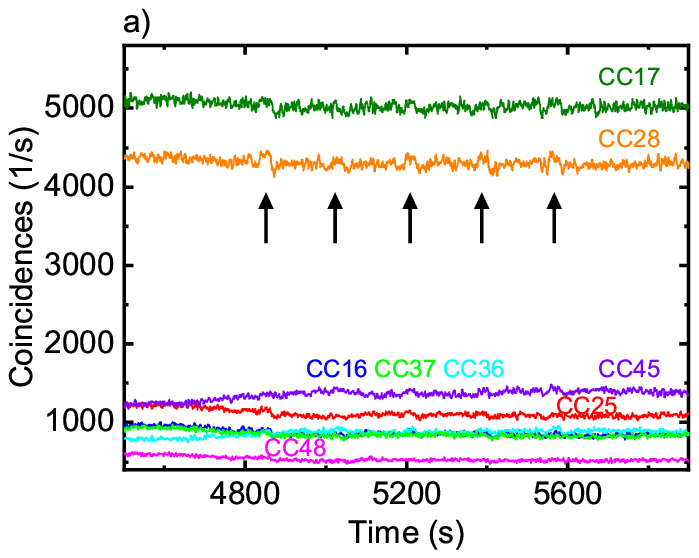}
\includegraphics[width=0.48\textwidth]{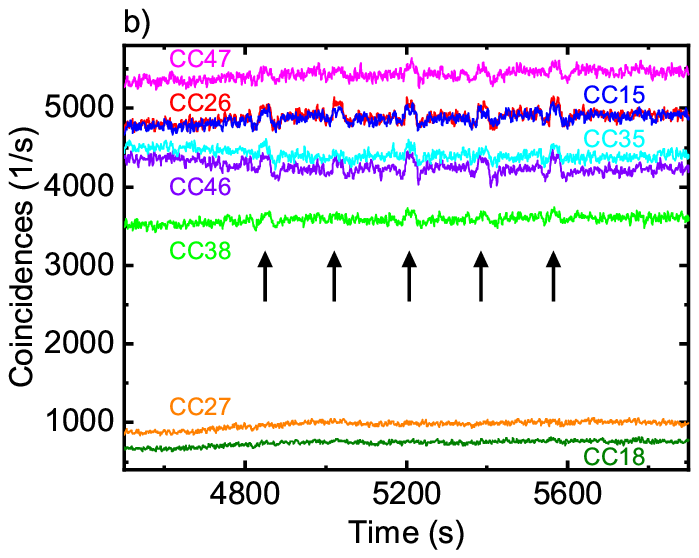}
\vspace*{-2mm}
\caption{\textbf{Coincidence rates}.
Moving average (with sliding window of 3~s) of the coincidence count rates for detector combinations that are used for the calculation of the $S$ value during parabolas 21 to 25 of the first flight. The plots are shown for the same time window as in Fig.~\ref{fig:Singles_Acc_over_time}.
Black arrows indicate time intervals at which the coincidence count rates of various detector pairs show qualitative correlations that with periods of significant acceleration changes.
As with the single-photon count rates of some detectors, the coincidence counts do not stay constant during the flight, but on short and long time scales, yet they give rise to almost constant $S$ values.
}
\label{fig:Coinc_over_time}
\end{figure*}


\subsection{Kolmogoroff-Smirnoff Test}\label{appendix:KST}

\noindent In order to statistically underpin the observation that the levels of acceleration (or changes thereof) during our experiment had no influence on the entanglement of the produced photon pairs we use
the two-sample Kolmogoroff-Smirnoff test (KST). In this hypothesis test, the empirical cumulative distribution functions (ECDFs) of two samples, e.g., taken during periods of steady flight and microgravity, respectively (see Fig.~\ref{fig:KST_ECDF_1-10_and_26-30}), are compared in order to estimate the likelihood that both samples are drawn from the same underlying distribution, for a review see, e.g., \cite{PrattGibbons1981}. Here, we use this test for pairwise comparisons of the ECDFs for both the $S$ values measured on the first day and the visibilities measured on the second day for pairs of samples corresponding to time intervals with different levels of acceleration. For these time intervals we selected periods of hypergravity (HG), microgravity (MG), and steady flight (SF) during which acceleration levels showed only small fluctuations (see Fig.~\ref{fig:KST_EDFs_HG_MG_SF_S}), as well as periods that we labelled "jerk" during which the acceleration changes but can be well approximated by a straight line whose slope well approximates the time derivative of the acceleration (the jerk), see Fig.~\ref{fig:KST_EDFs_Jerk_MG_S_and_Acc_vs_time}.\\[-2mm]

The null hypothesis $H_0$ that the KSTs we perform aim to check is that the different levels and/or changes of accelerations during the pairs of selected periods have no influence on the entanglement (as measured by the distribution of $S$ values and visibilities). For sufficiently large sample sizes $n_{1}$ and $n_{2}$, the null hypothesis is rejected at significance level $\alpha$ if the maximal distance
\begin{align}
d_{n_{1},n_{2}} &= \sup_{x} \vert F_{n_{1}}(x) - F_{n_{2}}(x) \vert
\end{align}
between the ECDFs $F_{n_{i}}(x)=\tfrac{1}{n_{i}}\sum_{j=1}^{n_{i}}\mathbf{1}_{x_{j}\leq x}$ (the fraction of samples with values less than or equal to $x$) with the indicator
\begin{align}
\mathbf{1}_{x_{j}\leq x} &= \begin{cases}
1   & \text{if}\ \ x_{j}\,\leq\,x\\
0   & \text{otherwise}
\end{cases}
\end{align}
is larger than a specific critical value $D_{\mathrm{c}}$, that is, if
\begin{align}
    d_{n_{1},n_{2}} &>\,D_{\mathrm{c}}(\alpha;n_{1},n_{2})\,=\,K_{\alpha}\ \sqrt{\frac{n_{1}\,+\,n_{2}}{n_{1}\,n_{2}}}\,=\,
    \sqrt{-\tfrac{1}{2}\ln\bigl(\tfrac{\alpha}{2}\bigr)}\ \sqrt{\frac{n_{1}\,+\,n_{2}}{n_{1}\,n_{2}}}\,.
\end{align}
In our case, we choose a significance level of 5~\%, which corresponds to setting $K_{\alpha = 0.05}=\sqrt{\ln(40)/2} = 1.3581$.\\[-2mm]

\begin{figure}[ht!]
\centering
\includegraphics[width=0.5\textwidth]{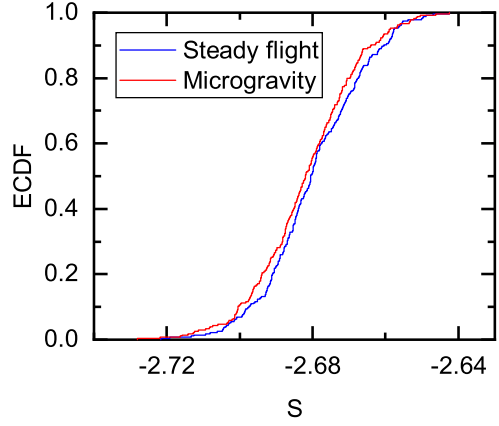}
\vspace*{-6mm}
\caption{\textbf{Empirical cumulative distribution functions (ECDFs) $F_{\mathrm{SF}}(S)$ and $F_{\mathrm{MG}}(S)$} for the $S$ values during steady flight (SF) and microgravity (MG) for a sample size of $n_1 = n_2 = 274$. The absolute values of the accelerations along the three spatial axes during steady flight have a mean of $0.98$~g and are in the range of $0.86\pm0.012$~g and $1.06\pm0.012$~g, for the microgravity phase the mean value of the absolute value of acceleration along all three spatial directions is $0.023$~g and the maximum and minimum accelerations are $0.085\pm0.012$~g and $0.002\pm0.012$~g, respectively. A Kolmogoroff-Smirnoff test provides the maximum difference of the ECDFs of $d_{\mathrm{max}} = 0.08$ at $S = -2.693$. With a significance level of 5\% the maximum difference is smaller than the critical value $D_{\mathrm{c}} = 0.12$. Thus both samples of $S$ values, corresponding to SF and MG, respectively, are consistent with the same underlying distribution, thus ruling out any influence of microgravity on the photonic entanglement in our setup with a 5~\% significance level. For this calculation and this plot we selected the microgravity $S$ values of parabolas 1 through to 10 and 26 through to 30.}
\label{fig:KST_ECDF_1-10_and_26-30}
\end{figure}

Figure~\ref{fig:KST_ECDF_1-10_and_26-30} shows the ECDFs $F_{\mathrm{SF}}(S)$ and $F_{\mathrm{MG}}(S)$ of $S$ values calculated from data recorded during 15 periods of steady flight (SF) and from the same number of periods of microgravity (MG), respectively.
The $S$ values from the steady-flight periods are taken from time windows shortly before or after the individual parabolic flight maneuvers. The sample sizes of both distributions are $n_1 = n_2 = 274$. Since the steady-flight periods during parabolas 11 to 25 feature relatively large fluctuations of the acceleration over time compared to the steady-flight periods between the other parabolas, we only considered the $S$ values of the parabolas 1 to 10 and 26 to 30 for this test.
The mean of the absolute values of acceleration during SF is 0.98~g and the maximum and minimum accelerations are 1.06~g and 0.86~g, respectively. During microgravity, the mean of the absolute value of acceleration amounts to 0.023~g. The largest measured (absolute value of) acceleration during microgravity was 0.085~g and the smallest was 0.002~g.
The maximum distance $d_{\mathrm{max}} = 0.08$ of the KST at $S = -2.6931$ lies below the critical value of $D_{\mathrm{c}} = 0.12$ for a significance level of 5~\%. Hence, the null hypothesis (no influence of gravity/acceleration on entanglement) is confirmed at the 5~\% significance level.\\[-2mm]

We further perform pairwise comparisons of the ECDFs of $S$ values during SF, MG, and HG, see Fig.~\ref{fig:KST_EDFs_HG_MG_SF_S} a). The $S$ values corresponding to HG and MG are taken from parabola 21 and the SF $S$ values are taken from a time window before the parabolic flight maneuver with small fluctuations of the acceleration. With a sample size of $n_{\mathrm{SF}} = n_{\mathrm{MG}} = n_{\mathrm{HG}} = 15$ the critical value is $D_{\mathrm{c}} = 0.47$ in all cases. The KST between the $S$ values of HG and SF yields a maximum distance of $d_{\mathrm{max}}$ = 0.2 and for the $S$ values of HG and MG yields $d_{\mathrm{max}}$~=~0.33. The null hypothesis is confirmed in both cases.\\[-3mm]

\begin{figure*}[h]
\centering
\includegraphics[width=0.9\textwidth]{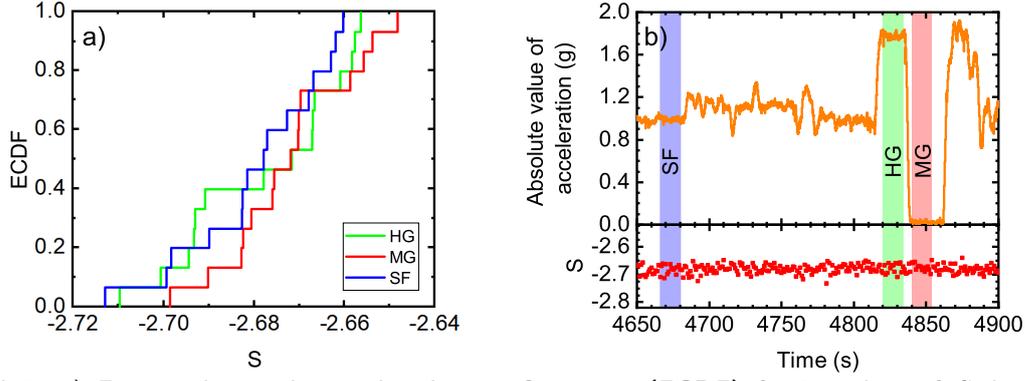}
\vspace*{-7mm}
\caption{a) \textbf{Empirical cumulative distribution functions (ECDF)} for 15 values of $S$ during hypergravity $F_{\mathrm{HG}}(S)$, microgravity $F_{\mathrm{MG}}(S)$, and steady flight $F_{\mathrm{SF}}(S)$.
b)
The $S$ values (red, bottom) and accelerations (orange, top) before and during parabola 21 are shown, with selected intervals of steady flight (SF), hypergravity (HG), and microgravity (MG) highlighted with blue, green, and red background colour, respectively.
The maximum distance between the ECDFs of the $S$ values during HG and SF is $d_{\mathrm{max}} = 0.2$ at $S = -2.662$, while the maximum distance between HG and MG is
$d_{\mathrm{max}} = 0.33$ at $S = -2.691$. Both are below the critical value of
$D_{\mathrm{c}} = 0.47$ (for a significance level of 5~\%) and thus the $S$ values of HG and SF, as well as those of HG and MG are confirmed to be consistent with the same distribution.
}
\label{fig:KST_EDFs_HG_MG_SF_S}
\end{figure*}

In this context it is also interesting to check whether the ECDFs of the $S$ values corresponding to the periods of MG and the periods of strongly changing accelerations (jerk) show a significant difference. For this purpose, we consider $S$ values from time periods in which the acceleration shows an approximately linear change as in Fig.~\ref{fig:KST_EDFs_Jerk_MG_S_and_Acc_vs_time}. For the ECDFs in Fig.~\ref{fig:KST_EDFs_Jerk_MG_S_and_Acc_vs_time}~a), we take the $S$ values from an 8-second time window of strongly changing acceleration and a preceding MG phase (sample size $n_{\mathrm{MG}} = n_{\mathrm{Jerk}} = 13$), indicated in b) by the purple and red background colour, respectively.
In this case, the null hypothesis is once again confirmed, with a maximum distance of $d_{\mathrm{max}} =0.38$ below the critical value of $D_{\mathrm{c}} = 0.46$. In the time window from $t = 2719$~s to $t = 2727$~s we perform a linear fit which yields a jerk of $j = -0.1357 \pm 0.0033$~g/s.\\[-3mm]

For twelve further jerk periods we perform the KST in the same way. For eleven tests the null hypothesis is confirmed. However, we found that in one case, the null hypothesis is not confirmed. If we take the $S$ values from $t = 2322$~s to 2329~s in migrogravity, and the $S$ values during an almost constant change of acceleration in the time window of $t = 2360$~s to $t=2367$~s (for which a linear fit provides a jerk of $-0.1346 \pm 0.0056$~g/s) we find $d_{\mathrm{max}} = 0.64$ while the critical value is $D_{\mathrm{c}} = 0.55$. We assume that this statistical outlier is caused by the small sample size $n_1 = n_2 = 11$.

Figure~\ref{fig:maxDistance_vs_Jerk} shows the results of the KSTs between the $S$ value distributions in microgravity and the $S$ value distributions from the periods of approximately constant change of acceleration.\\[-3mm]

To perform KSTs for the visibilities measured on the 27th of October we proceed in the same way as shown in Fig.~\ref{fig:KST_ECDF_1-10_and_26-30}. Figure~\ref{fig:KST_EDFS_VIS_HV_DA_SF_MG} shows the ECDFs of the measured visibilities in the H/V and D/A bases during SF and MG. For the ECDFs the maximum distance is $d_{\mathrm{max}} = 0.049$ for both the visibilities in the H/V basis as well as in the D/A basis. For the 5~\% significance level the critical value is $D_{\mathrm{c}} = 0.093$. Thus the null hypothesis is confirmed also for the visibilities.

\begin{figure*}[h!]
\centering
\includegraphics[width=0.9\textwidth]{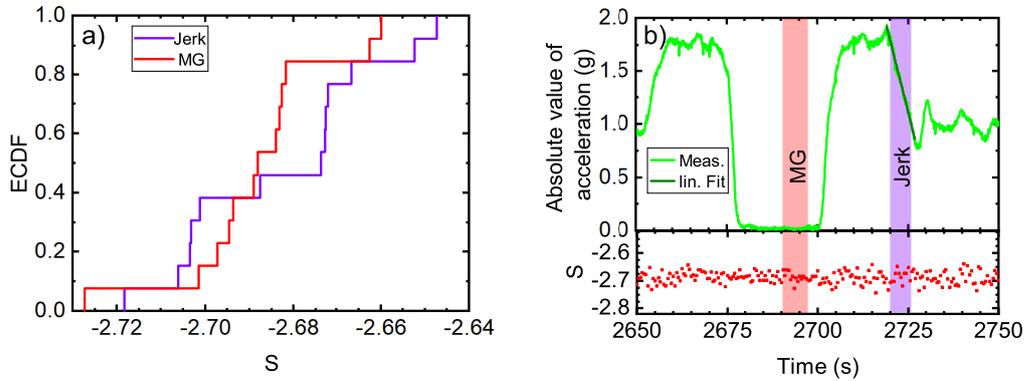}
\vspace*{-7mm}
\caption{a) \textbf{Empirical distribution functions (ECDF)} $F_{\mathrm{Jerk}}(S)$ and $F_{\mathrm{MG}}(S)$ for 13 values of $S$ recorded during periods of strongly changing acceleration (jerk periods) and microgravity (MG), respectively. For this KST, $S$ values were calculated with an integration time of 500~ms.
b) The $S$ values (red, bottom) and accelerations (green, top) during parabola 13 (flight on 26th of October) are shown.
In the time window between 2719~s and 2727~s the acceleration shows an almost linear change with time. A linear fit in this time window provides a jerk of $j = -0.1357 \pm 0.0033$~g/s. We perform a KST with $S$ values from this time window (purple background) and $S$ values taken from the adjacent microgravity phase (red background).
For a significance level of 5~\% and a sample size of 13 we get a critical value of $D_{\mathrm{c}} = 0.46$. With a maximum distance of $d_{\mathrm{max}} =0.38$ between the ECDFs at $S =-2.68178$ the critical value is not reached. Hence the $S$ values from MG and the constant change of acceleration are consistent with the same underlying distribution, which in turn confirms the null hypothesis that the measured jerk has no influence on our entangled photon pairs.
}
\label{fig:KST_EDFs_Jerk_MG_S_and_Acc_vs_time}
\end{figure*}

\begin{figure*}[h!]
\centering
\includegraphics[scale=1]{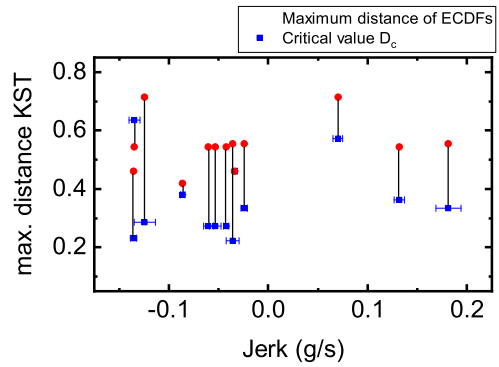}
\caption{\textbf{Kolmogorov–Smirnov tests comparing steady flight and strongly changing acceleration}. The maximum distances (blue dots, bottom, except for the topmost on the left-hand side) and their corresponding critical values $D_{\mathrm{c}}$ (red dots, top, except for the topmost on the left-hand side) for the KSTs between $S$ values in microgravity and $S$ values during an approximately constant change of acceleration (Jerk). Black bars connect the maximum distances with their corresponding critical values. Each red and blue dot pair is a single KST. The sample sizes of the KSTs are between 6 and 20. Error bars correspond to the 95~\% confidence interval of the linear fit on the acceleration. The maximum distance is below the critical value for all KSTs in this plot except for one: for the pair of ECDFs with a jerk of $j = -0.1346 \pm 0.0056$~g/s the maximum distance lies above the critical value. The sample size for this KST is 11.
}
\label{fig:maxDistance_vs_Jerk}
\end{figure*}

\begin{figure*}[h!]
\centering
\includegraphics[scale=1]{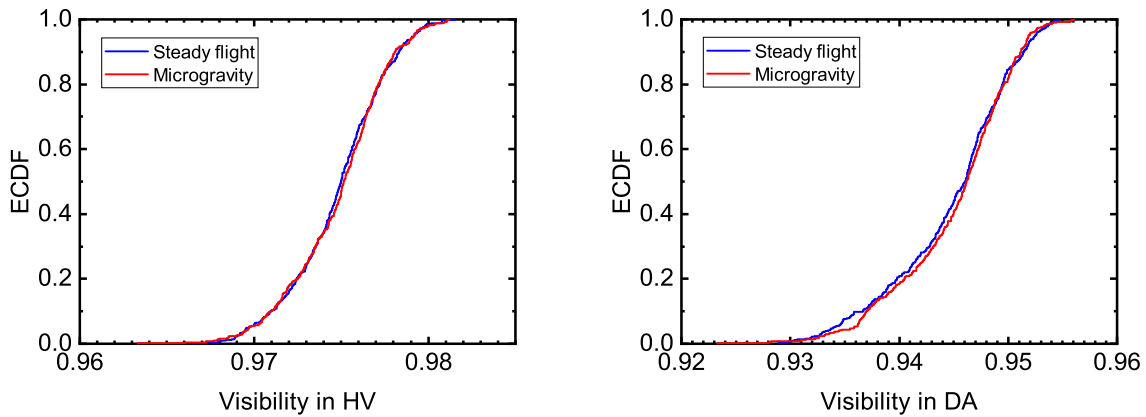}
\caption{\textbf{Empirical cumulative distribution functions (ECDF)} $F_{\mathrm{SF}}(V)$ and $F_{\mathrm{MG}}(V)$ of the visibilities in the H/V basis (left-hand side) and the D/A basis (right-hand side) during steady flight (SF, blue) and during microgravity (MG, red) (data from the 2nd flight on October 27th 2021).
For a sample size of 431 and a significance level of 5~\% the critical value for the KST is 0.093. For both the ECDFs of the visibility in H/V and in D/A the maximum distance is $d_{\mathrm{max}} = 0.049$ and thus below the aforementioned threshold of 0.093. Therefore, the visibility values of the SF and MG periods are consistent with the same underlying distribution with a 5~\% significance level.
The mean of the absolute values of the accelerations during steady flight is 0.98~g with a maximum of 1.11~g and a minimum of 0.87~g.
For the absolute values of the accelerations during microgravity the mean evaluates to 0.018~g, with a minimum of $0\pm0.012$~g and a maximum of $0.059\pm0.012$~g.
}
\label{fig:KST_EDFS_VIS_HV_DA_SF_MG}
\end{figure*}


\clearpage

\subsection{Data for all Parabolas}\label{appendix:parabola data}

In this final appendix, we present the entire data for all $S$ values and accelerations measured during the first flight (26th October 2021) in Figs.~\ref{fig:All_S_values_over_complete_flight} and~\ref{fig:All_S_values_over_complete_flight_6_rows}, as well as for all visibility measurements and accelerations measured during the second flight (27th October 2021) in Figs.~\ref{fig:All_Visibilities_values_over_complete_flight} and~\ref{fig:All_Visibilities_over_complete_flight_6_rows}.
\begin{figure*}[hb!]
\centering
\includegraphics[width=0.9\textwidth]{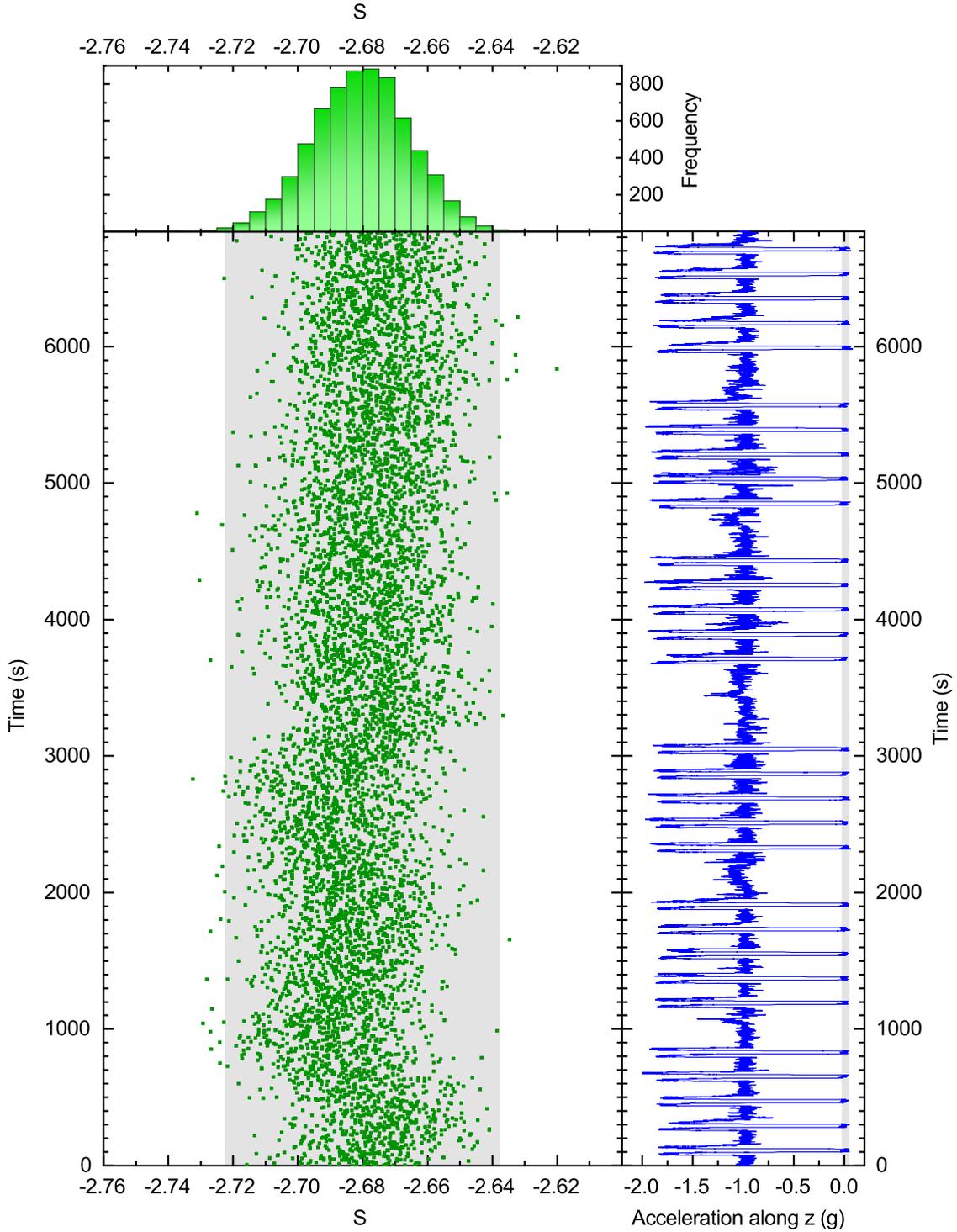}
\vspace*{-2mm}
\caption{\textbf{$S$ values and accelerations during first flight}. The plot shows all measured $S$ values during 30 parabolas of the measurement of flight one (green dots). The gray area highlights the area of $\pm 3$ standard deviations around the mean value of $S = -2.680$. The top plot shows a histogram of the measured $S$ values over the whole flight.
The measurement over 30 parabolic flight maneuvers lasted around $\thicksim$ 1.9~h.}
\label{fig:All_S_values_over_complete_flight}
\end{figure*}
\begin{figure*}[h!]
\centering
\includegraphics[width=0.95\textwidth]{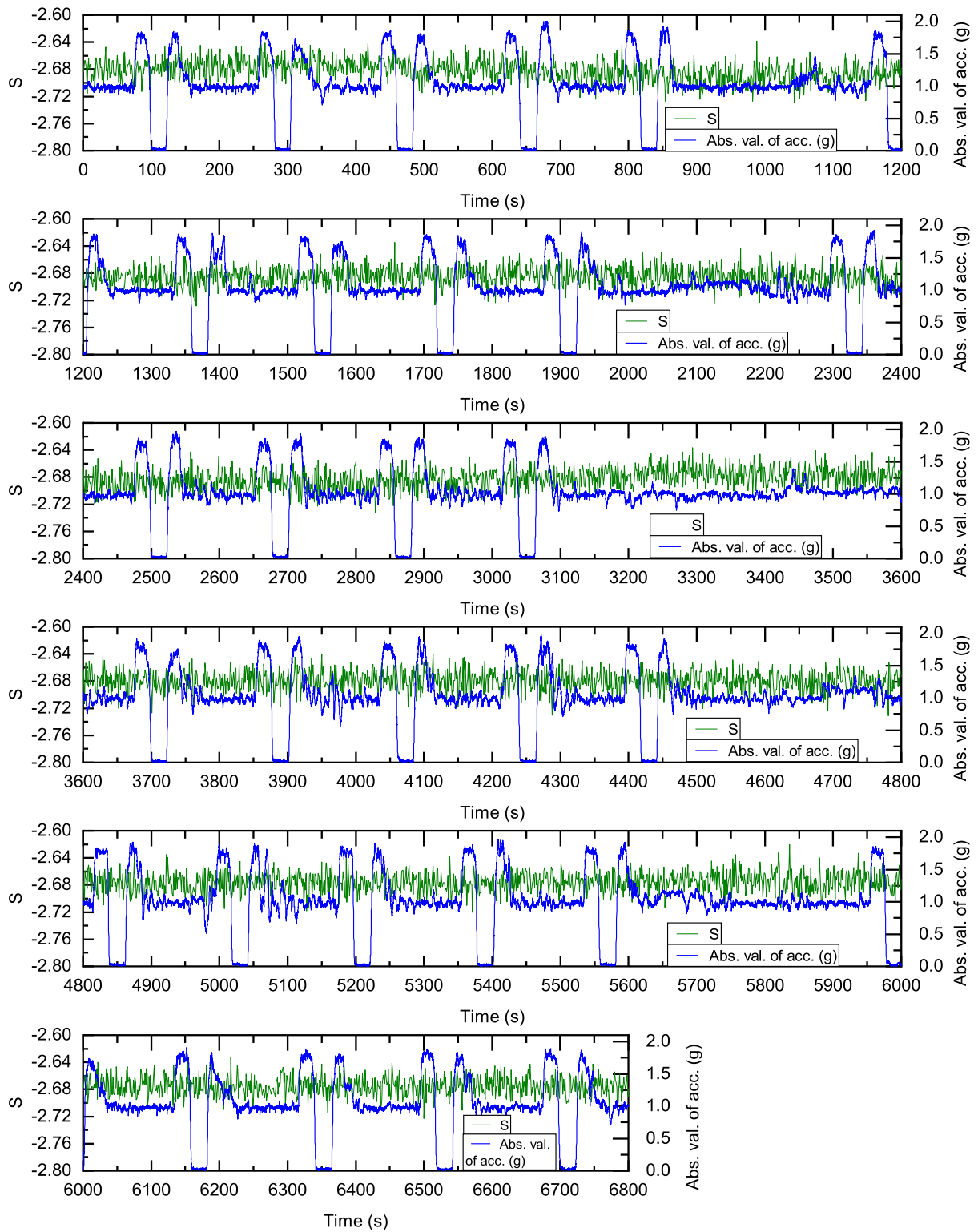}
\caption{\textbf{Overlay of $S$ values and accelerations during first flight}. The plot shows the measured $S$ values over 30 parabolas on flight one (green line) overlaid with the measured absolute values of acceleration (blue line).}
\label{fig:All_S_values_over_complete_flight_6_rows}
\end{figure*}

\begin{figure*}[h]
\centering
\includegraphics[width=1.0\textwidth]{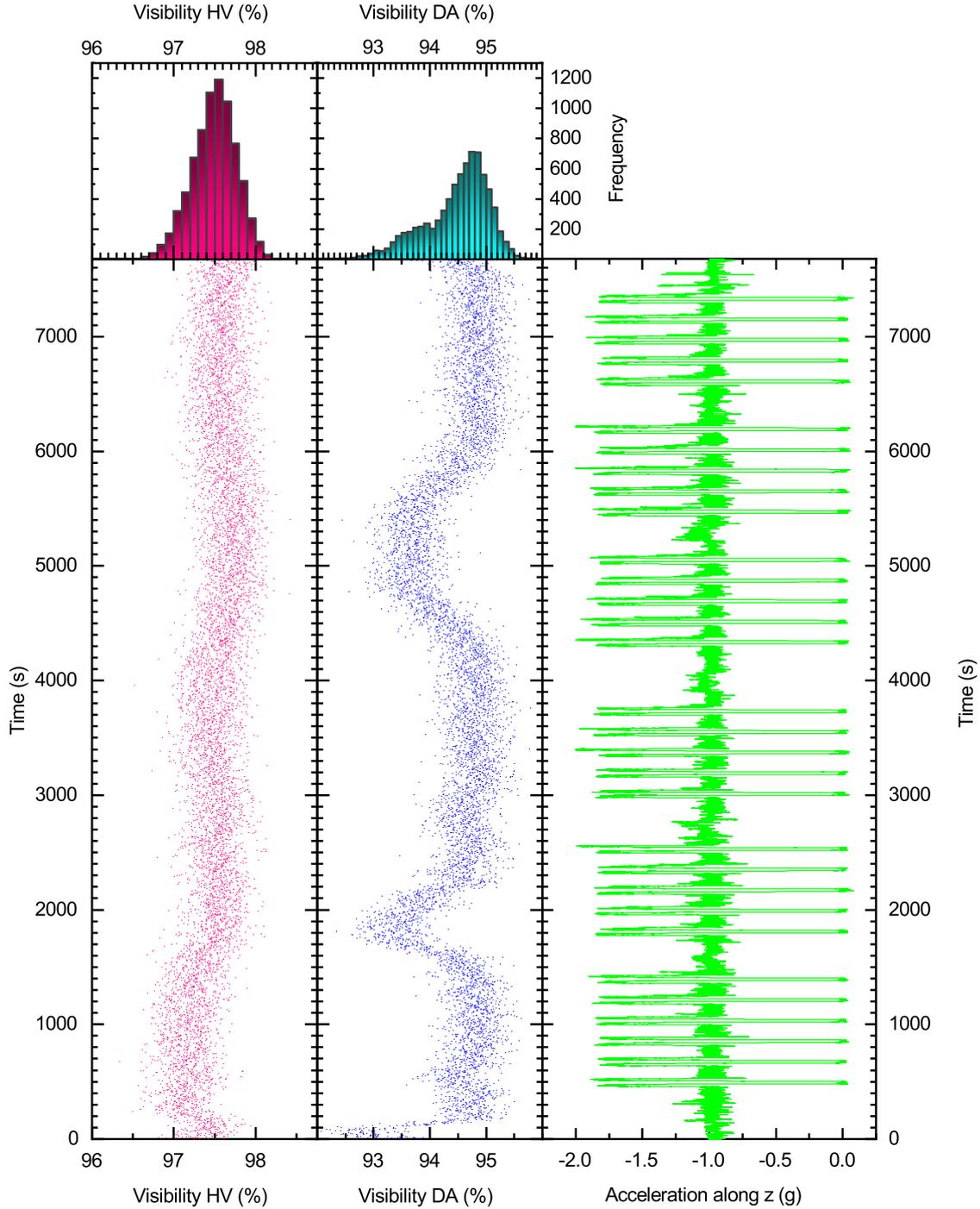}
\vspace*{-6mm}
\caption{
\textbf{Visibilities and accelerations during second flight}. The plot shows the visibilities in the H/V basis (magenta, left) and D/A basis (blue, middle) as well as the accelerations (green, right) measured during the 31 parabolas of flight two.
During the flight, the visibility in the H/V basis slightly increased over time while the visibility in the D/A basis drops twice to values around 92.33~\% ($t = 1813 \, \mathrm{s}$) and 92.44~\% ($t = 5398 \, \mathrm{s}$).
We found no correlation between these dips in visibility and any of the other measured parameters, in particular, the crystal temperature,
the cabin pressure, or the laser current (see Fig.~\ref{fig:Cryst_Temp_Pressure_Cabin_Temp}). Neither do these drops correlate with the acceleration (nor with the change of the acceleration). Since we measure a temperature increase over all parabolas
of about $\sim6.83^\circ \mathrm{C}$ $ (\sim7.79^\circ \mathrm{C})$ (see Fig.~\ref{fig:Cryst_Temp_Pressure_Cabin_Temp}) in the second level of the rack, we assume that the dips may be caused by heating of optical components.
There is also a dip in the visibility of the D/A basis before
the parabolic flight maneuvers start between $0 \, \mathrm{s}$ and $200 \,\mathrm{s}$. During this period we adjusted
the temperature of the crystal.
The whole measurement lasted around $\sim$ 2.1~h.
}
\label{fig:All_Visibilities_values_over_complete_flight}
\end{figure*}

\begin{figure*}[h]
\centering
\includegraphics{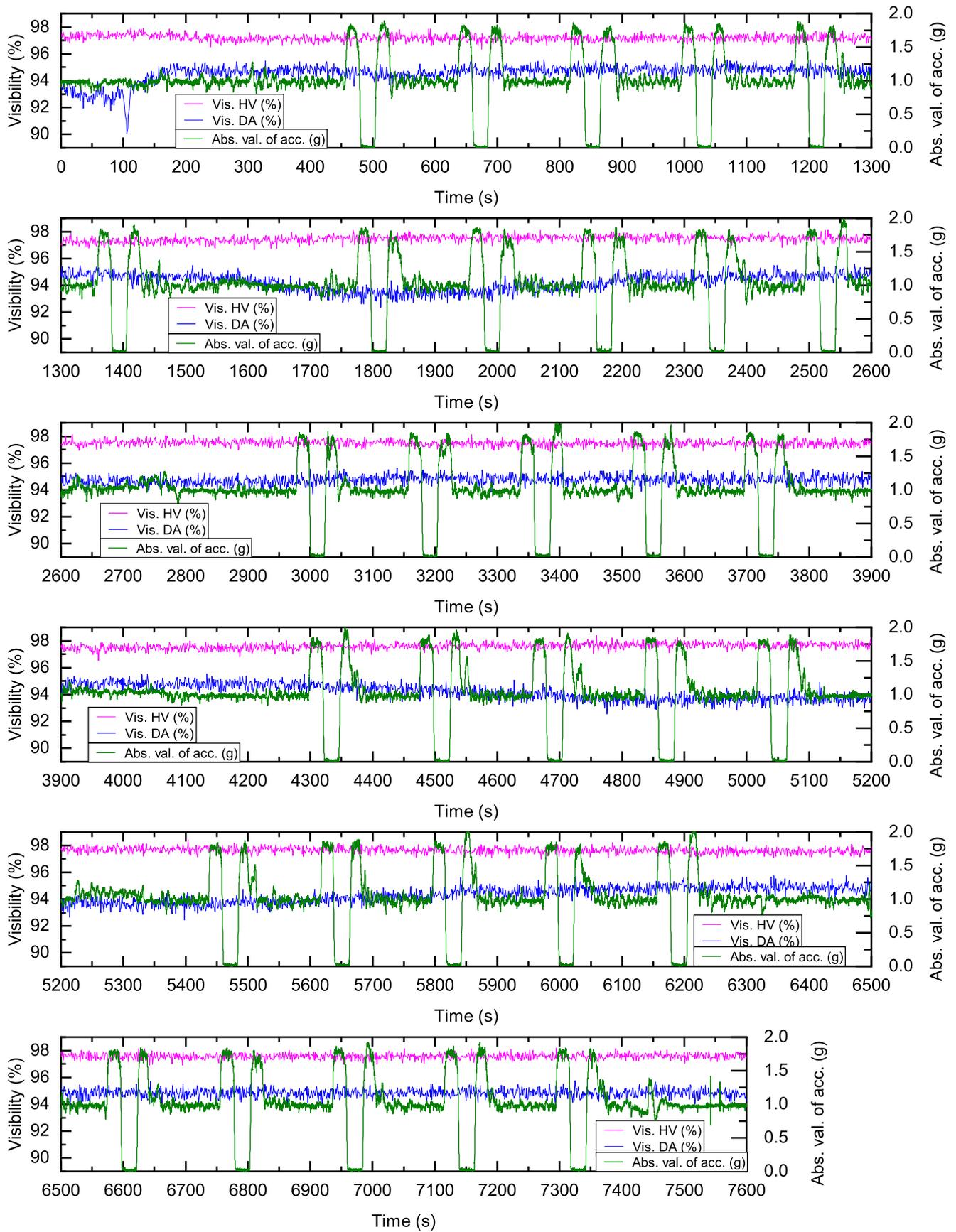}
\vspace*{-5mm}
\caption{
\textbf{Overlay of visibilities and accelerations during second flight}. The plot shows the measured visibilities in the H/V (magenta) and D/A (blue) bases over 31 parabolas on flight two overlaid with the measured absolute values of acceleration (blue line).
The very first dip of the visibility in the D/A basis (between $t = 50 \, \mathrm{s}$ and $t = 127 \, \mathrm{s}$) is due to our temperature adjustment before the start of the parabolic flight maneuvers.
}
\label{fig:All_Visibilities_over_complete_flight_6_rows}
\end{figure*}

\end{widetext}

\end{document}